\documentclass[10pt, conference, letterpaper]{IEEEtran}

\ifCLASSINFOpdf
\else
\fi

\usepackage{comment}
\usepackage{soul}
\usepackage{xcolor}
\usepackage{subfigure}
\usepackage{amssymb}
\usepackage{amsmath}
\usepackage{amsfonts}
\usepackage{graphicx}
\usepackage{bm}
\usepackage{xfrac}
\usepackage[acronym,shortcuts]{glossaries}
\usepackage{balance}
\usepackage{caption}
\usepackage{bbm}

\newacronym{pso}{PSO}{Particle Swarm Optimization}
\newacronym{ris}{RIS}{Reflective Intelligent Surface}
\newacronym{sinr}{SINR}{Signal-to-Interference Plus Noise Ratio}
\newacronym{snr}{SNR}{Signal-to-Noise Ratio}
\newacronym{vlc}{VLC}{Visible Light Communication}
\newacronym{nlos}{NLOS}{Non-Line-Of-Sight}
\newacronym{los}{LOS}{Line-Of-Sight}
\newacronym{pls}{PLS}{Physical Layer Security}
\newacronym{rf}{RF}{Radio-Frequency}
\newacronym{led}{LED}{Light Emitting Diodes}
\newacronym{uav}{UAV}{Unmanned Aerial Vehicles}
\newacronym{wbplsec}{WBPLSec}{Watermark Blind Physical Layer Security}
\newacronym{fov}{FOV}{Field-Of-View}
\newacronym{wdm}{WDM}{Wavelength Division Multiplexing}
\newacronym{ss}{SS}{Spread-Spectrum}
\newacronym{ask}{ASK}{Amplitude Shift Keying}
\newacronym{pd}{PD}{Photo-Diode}
\newacronym{im}{IM}{Intensity Modulation}
\newacronym{dc}{DC}{Direct-Current}
\newacronym{dd}{DD}{Direct-Detection}

\usepackage{amsthm}
\theoremstyle{remark}

\usepackage{cuted}

\usepackage{booktabs}

\usepackage{tabularx}
\usepackage{threeparttable} 
\usepackage{multirow} 
\usepackage{rotating} 
\usepackage{bigstrut} 

\newcommand{\parag}[1]{\noindent\textbf{#1. }}

\newcommand{\simone}[1]{{\textcolor{violet}{#1}}}

\begin{document}

\title{VLC Physical Layer Security through  RIS-aided Jamming Receiver for 6G Wireless Networks}

\author{\IEEEauthorblockN{Simone Soderi\IEEEauthorrefmark{1}~\IEEEmembership{Senior Member,~IEEE,},
Alessandro Brighente\IEEEauthorrefmark{2},
Federico Turrin\IEEEauthorrefmark{2}, 
and~Mauro Conti\IEEEauthorrefmark{2},~\IEEEmembership{Fellow,~IEEE}}
\IEEEauthorblockA{\IEEEauthorrefmark{1}IMT School for Advanced Studies, Lucca, Italy\\
 simone.soderi@imtlucca.it}
  \IEEEauthorblockA{
 \IEEEauthorrefmark{2}University of Padua, Department of Mathematics, Padua, Italy \\
 alessandro.brighente@unipd.it, \{conti, turrin\}@math.unipd.it}}


\maketitle

\begin{abstract}
\ac{vlc} is one the most promising enabling technology for future 6G networks to overcome~\ac{rf}-based communication limitations thanks to a broader bandwidth, higher data rate, and greater efficiency. However, from the security perspective, \acp{vlc}  suffer from all known wireless communication security threats (e.g., eavesdropping and integrity attacks).
For this reason, security researchers are proposing innovative \ac{pls} solutions to protect such communication. 
Among the different solutions, the novel~\ac{ris} technology coupled with~\acp{vlc} has been successfully demonstrated in recent work to improve the~\ac{vlc} communication capacity.
However, to date, the literature still lacks analysis and solutions to show the \ac{pls} capability of~\ac{ris}-based~\ac{vlc} communication.

In this paper, we combine watermarking and jamming primitives through the \ac{wbplsec} algorithm to secure \ac{vlc} communication at the physical layer. Our solution leverages \ac{ris} technology to improve the security properties of the communication.
By using an optimization framework, we can calculate \ac{ris} phases to maximize the \ac{wbplsec} jamming interference schema over a predefined area in the room. In particular, compared to a scenario without \ac{ris}, our solution improves the performance in terms of secrecy capacity without any assumption about the adversary's location. 
We validate through numerical evaluations the positive impact of \ac{ris}-aided solution to increase the secrecy capacity of the legitimate jamming receiver in a \ac{vlc} indoor scenario. Our results show that the introduction of \ac{ris} technology extends the area where secure communication occurs and that by increasing the number of \ac{ris} elements the outage probability decreases.

\end{abstract}

\begin{IEEEkeywords}
RIS, Physical Layer Security, VLC, Jamming, Watermarking, 6G.
\end{IEEEkeywords}

\section{Introduction}\label{sec:intro}

The amount of traffic data in 2021 is estimated at around 235.7 Exabytes per month, while in 2016 was estimated at around 73.1 Exabytes per month~\cite{web:ciscoreport}. To manage this great amount of data, the traditional~\ac{rf} communication is inefficient and unable to satisfy the high-demand~\cite{tsonev2015towards}. For this reason, the scientific community is exploring innovative technology to manage traffic demand. Recent technologies include the most famous 5G and its possible successor, 6G.
In this context,~\ac{vlc} is a communication medium that obtained great attention as enabling technology for 5G and beyond. Currently we can find application of~\acp{vlc} in many different sectors, spacing from smart home devices~\cite{tiwari2015smart}, vehicle-to-vehicle communication~\cite{dahri2019experimental} and underwater communication~\cite{acar2017comparing}. 

Differently from~\ac{rf}, the potential of~\ac{vlc} is that it offers an higher data rate~\cite{ndjiongue1999visible}, high speed and robustness against interference~\cite{jovicic2013visible}, a large available frequency spectrum~\cite{rajagopal2012ieee}, and a low cost implementation thanks to~\acp{led}~\cite{tanaka2000wireless}. Furthermore,~\ac{vlc} operates within the spectrum of visible light and uses light for both illumination and data transmission.
However,~\acp{vlc} present different challenges to achieving an optimal transmission. Signal loss and poor detection are often observed when transferring data at long distances and \ac{nlos}. Moreover, this can also occur when transferring data at short ranges with line-of-sight when the photodetector is not angled toward the incident light. To capture and steer the~\ac{vlc}, generally, receivers are equipped with a convex lens that cannot dynamically steer the incident light beam. This helps the receiver reduce signal loss and improve detection performance. Modern receivers are equipped with a mechanism to dynamically steer the incoming beam, such as stretching a flexible matrix containing dielectric nano-resonators or ultra-fast switching of mie-resonant silicon nanostructure. However, all these methods have a weak effect on the refracted beam, even at high optical intensity~\cite{komar2017electrically}.

To increase the quality of the received beam, recent works are investigating the implementation of~\ac{ris} element~\cite{9474926}.~\ac{ris} is a novel technology that implements electronically configurable physical characteristics, which can be obtained by varying the temperature of a nano-cell or re-orientating Liquid-Crystal alignment based on induced electrical field. The configurable characteristic of such technology allows obtaining great advantages in a wide range of applications. Proposals of~\ac{ris} application include unmanned aerial vehicles~\cite{mursia2021rise} and autonomous vehicles~\cite{ozcan2021reconfigurable}.

\parag{Motivation} The use of~\acp{vlc} in an indoor environment is considered safer than radio-based technology~\cite{BLINOWSKI2019246}. Due to the directivity of the optical signals and their high obstacles' impermeability, it is harder to intercept them from the outside. Therefore, it is less prone to the typical threats of wireless communication such as jamming or eavesdropping. Furthermore, the~\ac{ris} permits extremely efficient~\acp{vlc} signal propagation redirection at very little power consumption since varactor diodes can apply a phase shift or absorb the incoming signal in a real-time reconfigurable manner.

Therefore, the advantages~\ac{ris} can overcome some of the limitations of~\ac{vlc}. Recent studies proved that~\ac{ris} can also improve the \ac{vlc} communication performance~\cite{Ndjiongue:21, 9474926} from different point of view.
However,~\ac{ris} is a novel technology, and its application in \ac{vlc} has only been proposed in some preliminary works, and it has been only initially explored. 

\parag{Contribution} 
In this work, we propose a communication framework to increase the security \ac{vlc} communication at the physical layer. To do this, we implement the \ac{wbplsec} algorithm in \ac{ris}-aided \ac{vlc} communication. \ac{wbplsec} has been successfully implemented in different type of communication, including \ac{vlc}~\cite{2021:soderi_pls_rgb_vlc}. \ac{wbplsec} combines watermarking and jamming primitives at the receiver to achieve different security properties, including confidentiality, integrity, and ant-replay. To maximize the jamming interference obtained with \ac{ris} phases, we design and present an optimization framework that does not make any assumption on the attacker position.
Our results highlight the positive impact of the \ac{ris} technology to increase the secrecy capacity.

We summarize the contribution of our paper as follows:
\begin{itemize}
    \item We present a mechanism that leverages \ac{wbplsec} algorithm to secure \ac{vlc} communication at the physical layer and improve the security properties through \ac{ris}.
    \item We validate our framework through numerical simulations, showing the positive impact of \ac{ris} in terms of the secrecy capacity even without any assumption about the adversary's location.
    \item We show how our mechanism allows achieving different security properties, including replay attack resistance, high confidentially, and message integrity, enabling the protection of several threats. 
\end{itemize}

\parag{Organization}
The remainder of the paper is organized as follows. Section~\ref{sec:background} briefly recalls the concepts useful for understanding the paper, while Section~\ref{sec:related} discusses the related works. Section~\ref{sec:sys_model} describes the system model considered. Then, Section~\ref{sec:opt} introduces the our optimization framework and Section~\ref{sec:eval} presents the results of the simulations. Section~\ref{sec:security} discusses the security properties we achieve with our mechanism.
Finally, Section~\ref{sec:conclusion} concludes the paper by discussing our findings and the current framework limitations.

\section{Background}\label{sec:background}

In this section we briefly introduce the concepts useful to understand the reminder of the paper. In particular in Section~\ref{subsec:vlc_back} we summarize the \ac{vlc} channel model, in Section~\ref{subsec:ris_back} we recall the \ac{ris} technology, and in Section~\ref{subsec:wbplsec_vlc} we detail of the \ac{wbplsec} algorithm.

\subsection{VLC Channel Model}\label{subsec:vlc_back}

\begin{figure}[t]
	\centering
	\includegraphics[width=0.95\columnwidth]{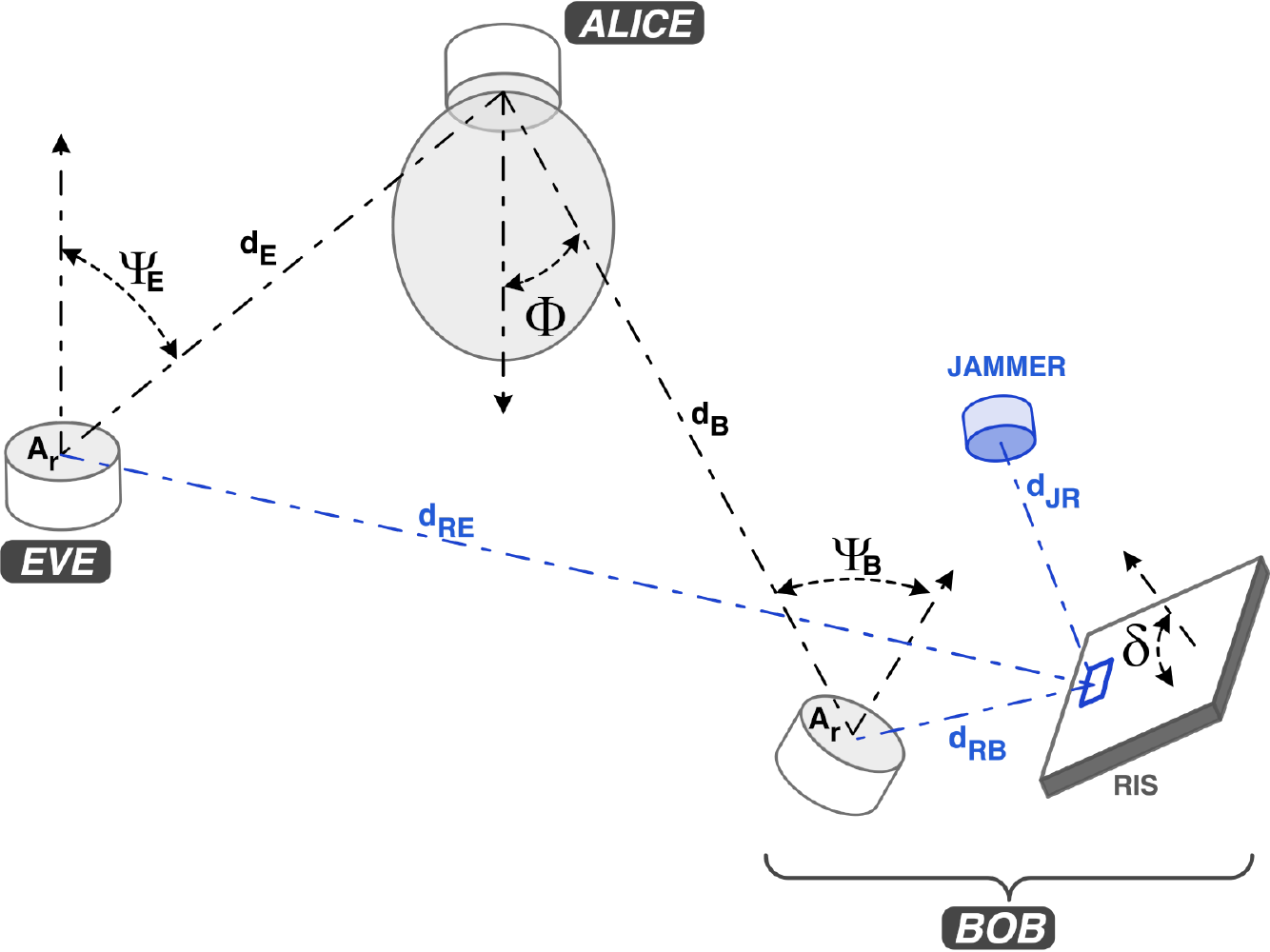}
	\caption{Contribution of \ac{ris} in the \ac{los} \ac{vlc} channel model. }
	\label{FIG:RIS-schema}
	\vspace{-1em}
\end{figure}
\acp{vlc} utilize the \ac{im} scheme along with \ac{dd}. The desired illumination level is maintained on the transmitter side by setting the appropriate \ac{dc} bias of the overall signal fed into the \ac{led}. On the other side, the received signal is proportional to the optical power that arrives at the photodiode.
An internal \ac{vlc} channel consists of two main components: the \ac{los} channel and the diffuse channel.
The first component considers the contribution of light directly hitting the photodiode without bouncing off other objects. The second component, also known as \ac{nlos}, includes all light rays that bounce off obstacles in the room. This article considers only the \ac{los} component (see Figure~\ref{FIG:RIS-schema}).
Assuming to have Lambertian light source, the \ac{los} component of the channel \ac{dc} gain, i.e. $H(0)~=~\int_{-\infty}^{\infty}h(t)dt$, between one \ac{led} and one photodiode is given by~\cite{1997:vlc_channel_barry}
%
%
\begin{equation}
\label{EQ:VLC_LOS}
H_{d}(0) = \begin{cases}
\frac{Ar(m+1) R }{2\pi d^2}  D(\psi)\cos^{m}(\phi) \cos(\psi), & |\psi| \leq \psi_{FOV}\\
0, & |\psi| > \psi_{FOV}
\end{cases},
\end{equation}

where $A_r$ is the receiver collection area, $R$ is the photodiode responsivity, $m = \sfrac{-\ln(2)}{\ln(\cos(\phi_{\frac{1}{2}}))}$ is the order of the Lambertian emission with half irradiance at $\phi_{\frac{1}{2}}$, $\phi$ is the angle of irradiance, $D(\psi) = \sfrac{n^2}{\sin^2(\psi_{FOV})}$ is the gain of the optical concentrator with $n$ refractive index, $d$ is the \ac{los} distance between the \ac{led} and the photodiode,  $\psi$ is the angle of incidence ($\psi \in \{\Psi_E,\Psi_B\}$), and $\psi_{FOV}$ is the receiver's angle \ac{fov}.

The total received power for Alice with one RGB \ac{led}, for a given transmission power ($P_t$), is given by the \ac{dc} channel on the directed path, i.e. $P_r = P_t \cdot H_{d}(0)$.

In a \ac{vlc} system, the \ac{snr}, i.e. $\gamma_{v}$, is proportional to the square of the received optical power, i.e., $\gamma_{v}=\sfrac{H_d^2(0) P^2_t}{\sigma^2}$ where $P_t$ is the transmitted optical power, $H_d(0)$ is the channel \ac{dc} gain and  $\sigma^2$ is the spectral density of the background noise.

In terms of reflections on the \ac{ris}, we can define the channel \ac{dc} gain of the first reflection as follows
%
%
\begin{equation}
\label{EQ:VLC_NLOS}
dH_{ref}(0)= \begin{cases}
 \frac{A_r(m+1)R}{2\pi^2 d_{1B}^2 d_{2B}^2} D(\psi) \rho dA_{w}\cos^{m}(\phi)\cdot \\ \quad\cdot \cos(\alpha) \cos(\beta) \cos(\psi) & |\psi| \leq \psi_{FOV}\\
0 & |\psi| > \psi_{FOV}
\end{cases},
\end{equation}
where $\rho$ denotes the reflection coefficient, $dA_{w}$ represents the emission area of a micro surface, $\alpha$ expresses the incidence angle of a reflection point and $\beta$ is the radiation angle of the receiver.

To simplify the mathematical treatment in the remainder of the paper we will denote by $h$ instead of $H_d(0)$, the coefficients of the \ac{vlc} channel.

\subsection{Reflective Intelligent Surfaces}\label{subsec:ris_back}
A \ac{ris} provides a re-configurable reflection channel that can be exploited, among the others, to suitably direct signals in predefined directions. We adopt the \ac{ris} model in~\cite{aboagye2021intelligent}, where each element of the \ac{ris} acts as a mirror reflecting light in a suitably selected direction. As \ac{vlc} highly depend on the existance of a \ac{los} channel, thanks to \ac{ris} it is possible to relax this condition. Furthermore, \ac{ris} can serve the same function a beamformer serves in classical \ac{rf} communications, hence creating suitably combined signals~\cite{aboagye2021intelligent}. The main advantage of \ac{ris} is that it is passive, hence minimizing the overall system power requirements. In this paper, we model the signal reflected by the \ac{ris} via the \ac{nlos} \ac{vlc} channel~\eqref{EQ:VLC_NLOS}.

\subsection{WBPLSec Standalone Security Solution}\label{subsec:wbplsec_vlc}

\ac{wbplsec} algorithm utilizes a jamming receiver in conjunction with a \ac{ss} watermarking technique. As a standalone security solution for sensor networks, this technology has received considerable attention in recent years.
\ac{wbplsec} has been successfully applied in different context, including wireless communication~\cite{2017:soderi_wbplsec_trans} and acoustic communication~\cite{2019:soderi_acoustic_journal}. With this technique, the legitimate receiver creates a \textit{security region} around itself by leveraging jamming, enabling communication with a high degree of confidentiality under certain conditions. More precisely, to make jamming effective, the communication channel must allow for intentionally interfering with data transmission. Communication channels with possible transmission collision are most likely to fit this condition, e.g., wireless networks.
Secondly, network receiver nodes must be equipped with at least one transmitter and one receiver, which can be used simultaneously.

\begin{figure}[t]
\centering
\subfigure[System model without \ac{ris}.]{\includegraphics[width=0.9\columnwidth]{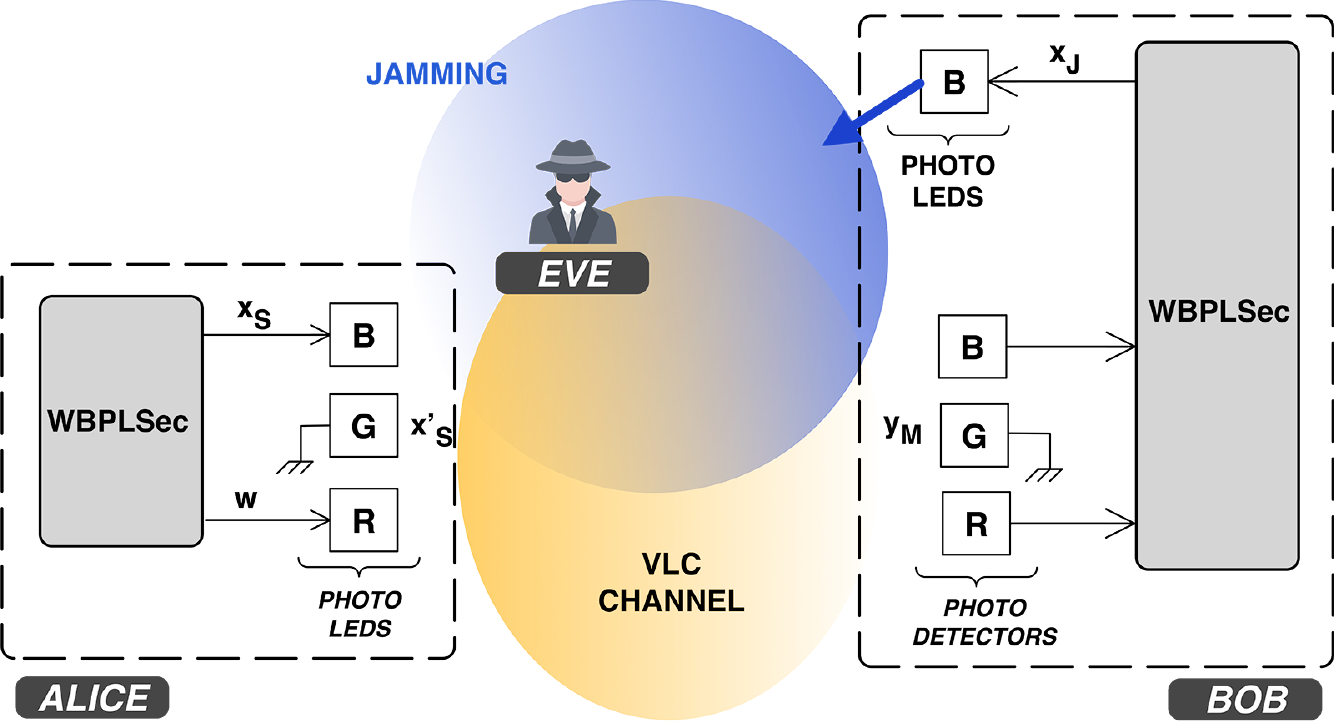}\label{subfig:SYS-NORIS}}\\
\subfigure[System model with \ac{ris}.]{\includegraphics[width=0.9\columnwidth]{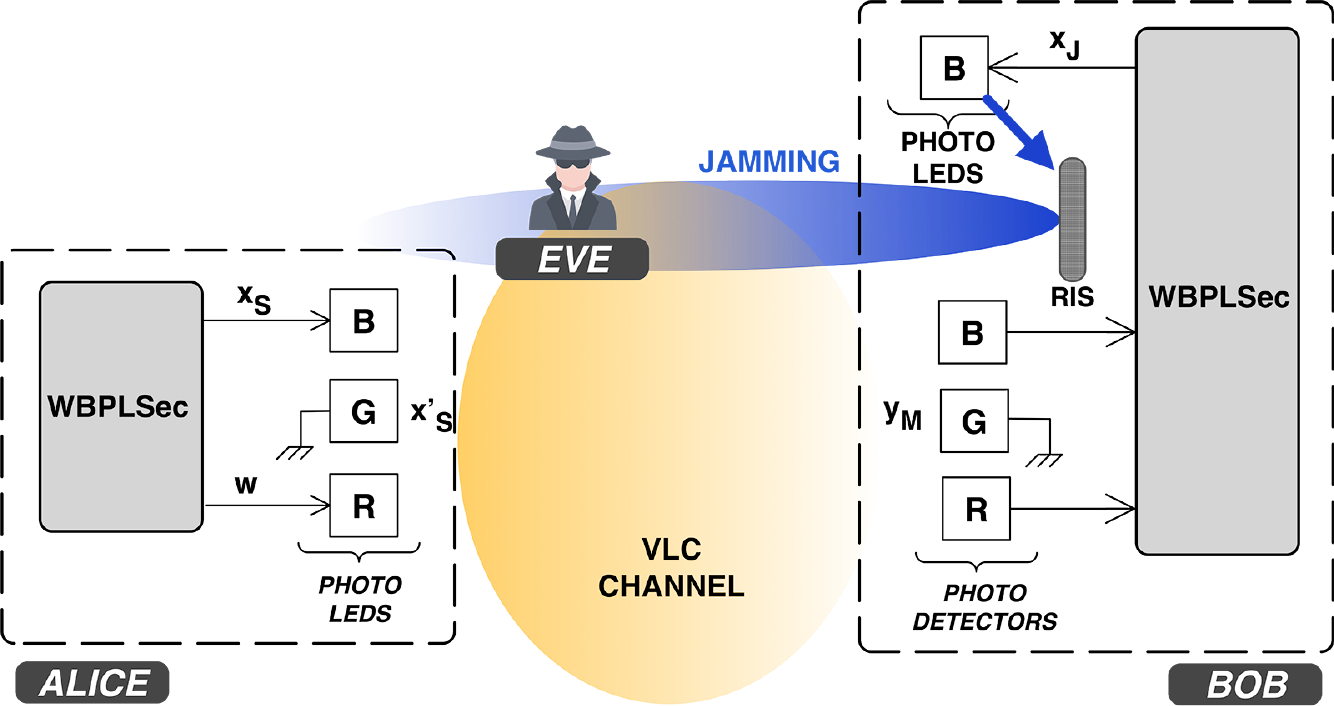}\label{subfig:SYS-RIS}}
\caption{System model using \ac{wbplsec} for \ac{vlc} in the two cases with and without \ac{ris}. } 
\label{FIG:SYS-RIS} 
\vspace{-1em}
\end{figure}

\parag{VLC Application Domain}
The \ac{wbplsec} protocol has been recently proposed to achieve confidentiality in \ac{vlc}~\cite{2021:soderi_pls_rgb_vlc}. In that case, watermarking and jamming are combined in \ac{wbplsec} to provide a standalone security solution in 6G networks.
That solution does not include the use of any \ac{ris}, which is the subject of this paper.
There are several ways to implement \ac{vlc}. The one that uses RGB \acp{led} provides the most bandwidth since it also uses three independent channels. Suppose an architecture that uses RGB \acp{led} to implement \ac{wbplsec} on \ac{vlc} depicted in Figure~\ref{subfig:SYS-NORIS}.

Let's consider the case that Alice wants to send a secret message of $N$ bits $(x_S)^N$ to Bob. Alice transmits the watermarked signal $(x_S')^N$ using an RGB \ac{led}. Bob receives the message through a single RGB color-tuned photodiode, but he jams $M$ ($M<N$) bits of the Alice’s \ac{vlc} $(x_J)^M$ using an RGB \ac{led} while receiving it. Eve, the eavesdropper, may use multiple PDs to violet secrecy.
The scheme proposed~\cite{2021:soderi_pls_rgb_vlc} (see Figure~\ref{subfig:SYS-NORIS}) exploits three RGB independent channels and uses a Wavelength Division Multiplexing (WDM) to watermark the \ac{vlc}. It relies on four main actions:
\textit{(i)} \textit{\ac{ss} watermarking}: part of the secret the message, i.e., $N_W$ of $N$ bits,  are first modulated with a spreading sequence to create the watermark signal $(w)^{N_W}$ and then transmitted by using only the red light;
\textit{(ii)} \textit{jamming receiver}: Bob jams Alice’s message using RGB \ac{led};
\textit{(iii)} \textit{selective jamming}: Bob jams only part of the on-the-way message, and he can rebuild the clean message by knowing the jammed part. The jamming does not affect the \ac{ss} watermark~\cite{2000:Proakis_digital_4th};
\textit{(iv)} \textit{communication hiding}: the proposed method transmits information through two independent paths using blue and red lights. The first is the narrow-band \ac{ask} signal transmitted through the blue light, i.e., $(x_S)^N$. At the same time, the \ac{ss} watermark signal uses the red light  $(w)^{N_W}$ and creates, in this way, a covert channel.

It is worth noting that compared to applying \ac{wbplsec} in \ac{rf} wireless communications, in the case of RGB \ac{led}-based \acp{vlc}, \ac{ss} watermarking and jamming primitives can be implemented without requiring additional transceivers.

\section{Related Works}\label{sec:related}

\ac{vlc} is a widely studied topic in literature. Different studies aim at implementing this type of communication in different fields, such as Vehicle-To-Vehicle networks~\cite{dahri2019experimental} or underwater communication~\cite{akram2017camera}. The wide diffusion and application of \ac{vlc} have raised the attention of the security community to investigate the vulnerabilities of such a medium. Different works investigate the physical layer security of \ac{vlc}~\cite{arfaoui2020physical} focusing on different properties. The different proposals to deal with \ac{pls} in \ac{vlc} include  beamforming~\cite{arfaoui2018secrecy, chen2017physical}, friendly jamming~\cite{wang2019enhancing, panayirci2020physical}, and signal mapping~\cite{yang2018mapping, jiang2017secrecy}. However, these works focus on the traditional implementation of \ac{vlc} communication.

Among the different enabling technologies used in the \ac{vlc} context, \ac{ris} represents a very recent and promising technology to improve the wireless communication performances~\cite{9474926}. In this paper we specifically focus on the security of \ac{vlc} in \ac{ris} application from the \ac{pls} point of view. 

Although different works propose the integration on \ac{ris} in different fields (e.g., \ac{uav}~\cite{mursia2021rise}), there is a lack of study on the security properties of this new technology.
The integration of \ac{ris} with attributes was studied in~\cite{abumarshoud2021lifi}, which particularly focus on the LiFi application of \ac{vlc}. Here the authors discussed how \ac{ris} can improve different \ac{pls} properties such as Secrecy capacity, jamming protection, and secure beamforming. However, the paper does not inspect the security topic deeply and does not provide secrecy capacity estimation. In~\cite{li2021robust} the authors propose a mechanism which implement \ac{ris} to improve the \acp{uav} communication security through alternating optimization technique. The paper also demonstrates that the proposed algorithm improves the average secrecy rate compared with other benchmark algorithms.

Differently from previous work we aim at implementing \ac{ris} and \ac{wbplsec}  protocol~\cite{2017:soderi_wbplsec_trans} to improve the \ac{pls} of \ac{vlc} communication (see Figure~\ref{subfig:SYS-RIS}). This technique was already applied on different communication means such as radio frequency ~\cite{2017:soderi_wbplsec_trans}, acoustic communications~\cite{2019:soderi_acoustic_journal} showing the improvement of performances from the secrecy capacity of the channel point of view. We show that the integration of \ac{ris} in \ac{vlc} communication allows \ac{wbplsec} to achieve even better performance than is a non \ac{ris}-aided \ac{vlc} communication~\cite{2021:soderi_pls_rgb_vlc}.
We then discuss the benefits of introducing \ac{wbplsec} in the communication and the security properties that it introduces in the communication.

\section{System Model}\label{sec:sys_model}

%
\begin{figure}[t]
	\centering
	\includegraphics[width=0.95\columnwidth]{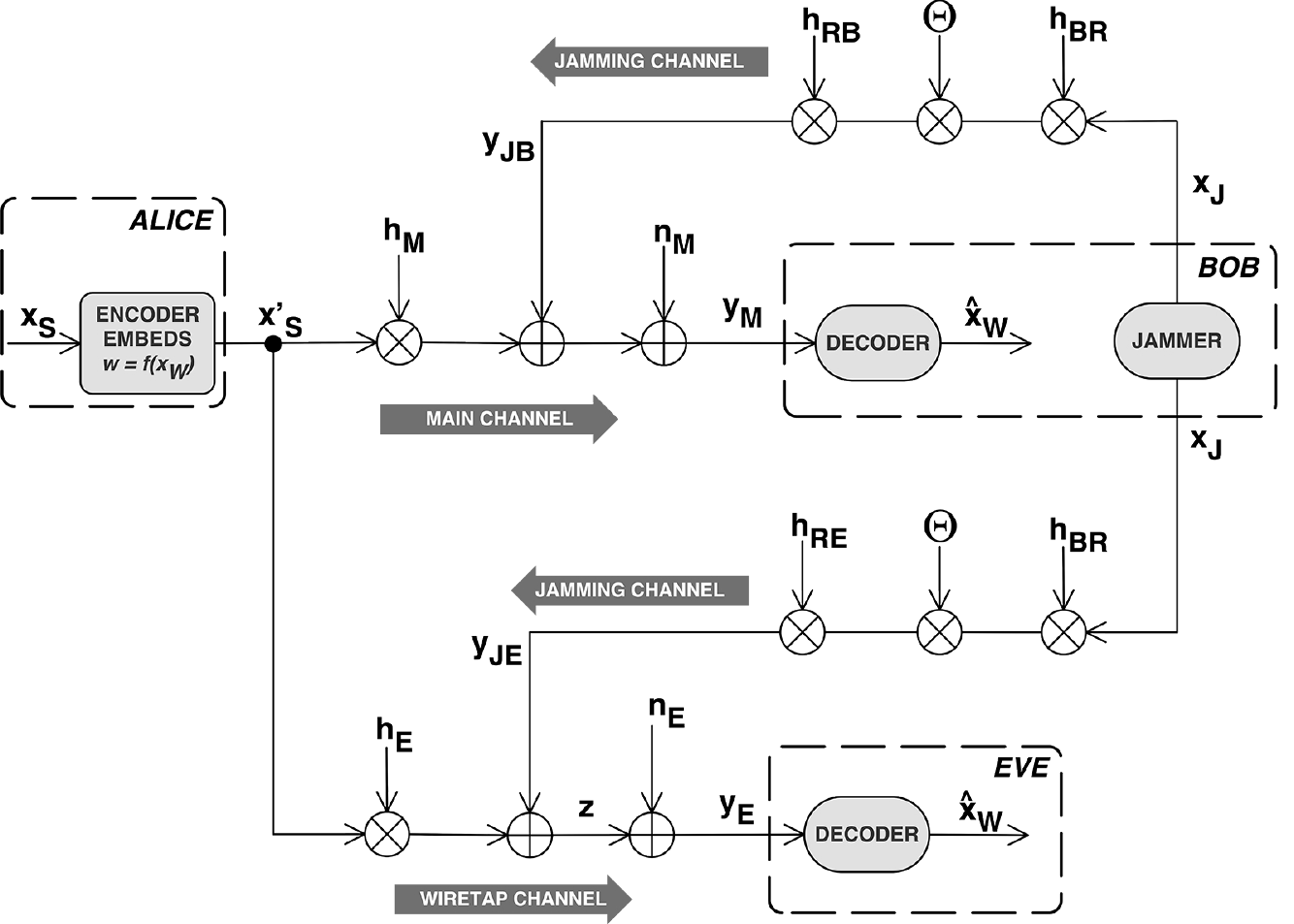}
	\caption{Modified non-degraded wiretap channel model with jamming receiver and \ac{ris} contribution. }
	\label{FIG:CH}
	\vspace{-1em}
\end{figure}
Let us consider the scenario depicted in Figure~\ref{FIG:CH}.
The legitimate user, \textit{Alice}, transmits $(x_S')^{N}$ to the legitimate receiver, \textit{Bob}, through the \emph{main channel}. A malicious eavesdropper, \textit{Eve}, receives this signal through the \emph{wiretap channel}. In our model, Bob is equipped with a jammer that generates a jamming signal to degrade the quality of the main channel by applying \ac{wbplsec}. Bob hence simultaneously receives Alice's signal and controls the jamming signal. We assume that the jammer transmits a signal $(x_J)^{M}$ and exploits a \ac{ris} to suitably select the jamming signal direction.
We denote as $\bm{h}_{JR}\in \mathbb{R}^{K \times 1}$ the channel between the jammer and the \ac{ris}. The reflected signal impacts both on the signal received by Bob and Eve. Let us denote as $\bm{h}_{RB} \in \mathbb{R}^{1 \times K}$ ad $\bm{h}_{RE} \in \mathbb{R}^{1 \times K}$ the channel from the \ac{ris} to Bob and from the \ac{ris} to Eve, respectively. The reflected channel is given by the \ac{nlos} channel~\eqref{EQ:VLC_NLOS}.
Therefore
the jamming signal at Bob can be computed as
\begin{equation}\label{eq:jammingBob}
    y_{JB}(\bm{\delta}) = \bm{h}_{RB}(\bm{\delta}) \bm{h}_{JR} x_j,
\end{equation}
whereas the received jamming signal at Eve as
\begin{equation}\label{eq:jammingEve}
    y_{JE}(\bm{\delta}) = \bm{h}_{RE}(\bm{\delta}) \bm{h}_{JR} x_j,
\end{equation}
where $\bm{\delta}=[\delta_1, \ldots, \delta_K]$ denotes the yaw angle of the each \ac{ris} element, as described in~\cite{aboagye2021intelligent}.


In accordance with the \ac{wbplsec} applied to VLCs~\cite{2021:soderi_pls_rgb_vlc}, considering both the useful signal and the jamming signal, the signal received by Bob and Eve are respectively:
%
%
\begin{align}
y_{M} &= h_{B} x_S'+ y_{JB}(\bm{\delta}) + n_{M}, \label{EQ:MainCH}\\
y_{E} &= h_{E} x_S' +  y_{JE}(\bm{\delta}) + n_{E}, \label{EQ:EveCH}
\end{align} 
where $h_{B}$ and $h_{E}$ are the channel's gains between Alice with Bob and Eve respectively, $x_S'$ is Alice's watermarked data signal, and $n_{B}$ and $n_{E}$ are the complex zero-mean Gaussian noise with variance $\sigma^2_E$ and $\sigma^2_B$, respectively. Notice that we neglected the noise terms in~\eqref{eq:jammingBob} and~\eqref{eq:jammingEve} to only consider it once in the overall received signals in~\eqref{EQ:MainCH} and~\eqref{EQ:EveCH}.

Given the received signals, we can define the \ac{sinr} at Bob's side as
%
%
\begin{equation}
    \label{EQ:SINR_M}
    \gamma_{M} = \frac{|h_B|^2 P_t^2}{\sigma_B^2 + |y_{JB}|^2 P_j^2}.
\end{equation}

The \ac{sinr} at Eve's side is given by
%
%
\begin{equation}
    \label{EQ:SINR_E}
    \gamma_{E} = \frac{|h_E|^2 P_t^2}{\sigma_E^2 + |y_{JE}|^2 P_j^2}.
\end{equation}

We assume that $\mathbb{E}[|x_S'|^2]=1$.
Furthermore, assuming that $\mathbb{E}[|x_j|^2]=1$, the interference component is given by
\begin{equation}
    \mathbb{E}[|y_{JB}|^2] = \mathbb{E}[|\bm{h}_{RE}\bm{\Theta} \bm{h}_{BR}|^2].
\end{equation}


\parag{Secrecy Capacity}

The idea we develop in this paper is to exploit the \ac{ris} configuration properties to improve the system performance given by \ac{wbplsec} in terms of information confidentiality. The secrecy capacity of the legitimate link for non-degraded Gaussian wiretap channels~\cite{1978:nonDegradedChannel, 2006:Barros_secrecy_wireless_ch} is a widely accepted metric for confidentiality at the physical layer. It can be defined as
%
%
%
\begin{equation}
	C_{s}= \max \{C_{M} - C_{E},0\}        
	= \begin{cases}
		\frac{1}{2} \log_2 \frac{1+\gamma_{M}}{1+\gamma_{E}}, \hspace{12pt} \text{if $\gamma_{M} > \gamma_{E}$}, \\
		0, \hspace{60pt} \text{if $\gamma_{M} \leq \gamma_{E}$}.
    \end{cases}
    \label{eq:SecCap}
\end{equation}
where $C_{M} = \log_2(1+\gamma_M)$ is the channel capacity from Alice to Bob, i.e. the main channel, and $C_{E}=\log_2(1+\gamma_E)$ is the channel capacity from Alice to Eve, i.e. the wiretap channel exploited by the eavesdropper.

\parag{Area Secrecy Capacity}

Although secrecy capacity provides a measure of confidentiality, in some scenarios, it relies on the limiting assumption that the location of Eve is known. To remove this assumption, we propose the use of \textit{area secrecy capacity}~\cite{uavRis}, which defines the average secrecy capacity over a predefined area. We denote as $\mathcal{A} \in \mathbb{R}^3$ the area of interest, and as $C_{E}(a), \gamma_E(a)$ as the channel capacity and the \ac{sinr} for an eavesdropper at location $a \in \mathcal{A}$. 
By denoting as $|\mathcal{A}|$ the area value of $\mathcal{A}$, we define the area secrecy capacity as
\begin{equation}
\begin{split}
    C_s(\mathcal{A}) = & \frac{1}{|\mathcal{A}|}\int_{\mathcal{A}}(\max \{C_{M} - C_{E}(a),0\})  da =   \\
    = & \begin{cases}
		\frac{1}{2|\mathcal{A}|}\int_{\mathcal{A}} \log_2 \frac{1+\gamma_{M}}{1+\gamma_{E}(a)} \, da, \hspace{30pt} \text{if $\gamma_{M} > \gamma_{E}(a)$}, \\
		0, \hspace{122pt} \text{if $\gamma_{M} \leq \gamma_{E}(a)$}.
    \end{cases}
    \end{split}
    \label{eq:areaSec}
\end{equation}
The area secrecy capacity provides a sub-optimal metric as, compared to the secrecy capacity in~\eqref{eq:SecCap}, it does not consider the actual Eve's channel. Instead, it exploits an average estimation of the punctual secrecy capacities obtained over hypothetical Eve's locations over an area of interest.  
However, by removing the assumption on Eve's location, this metric provides high generalization and applicability to those scenarios where the estimation of Eve's location is challenging.


\section{RIS-Adided Jamming Optimization}\label{sec:opt}

In this section, we define the optimization problem to obtain the \ac{ris} phases configurations that maximizes the communication secrecy. To show the advantages obtained using the area secrecy capacity to avoid assuming a known eavesdropper location, we provide the solution of two different problems: i) optimization with known Eve location (Section~\ref{sec:known}), and ii) optimization with unknown Eve location (Section~\ref{sec:unknown}).

\subsection{Known Eve Location}\label{sec:known}
Assuming a known Eve location, we achieve the best communication secrecy by maximizing the secrecy capacity~\eqref{eq:SecCap},  through the adjustment of RIS phases. Therefore, we define the following problem
\begin{subequations}\label{prob:overall}
\begin{equation}
    \max_{\bm{\Theta}} C_s,
\end{equation}
\begin{equation}
    \text{s.t.} \, \theta_k \in [0, 2 \pi], \, \forall k=1, \cdots, K.
\end{equation}
\end{subequations}
Notice that we only optimize the phases of the \ac{ris} as we are dealing with a single-receiver scenario. Therefore the optimal value of $C_s$ is achieved via maximal power transmission. Regarding the jamming power $P_j$, the best results would be achieved by choosing the maximum available power. However, we investigate via numerical evaluation the impact of $P_j$.

Problem~\eqref{prob:overall} is non-convex~\cite{boyd2004convex}, and finding the optimal solutions is a complicated task. However, we notice\simone{d} that, thanks to the application of \ac{wbplsec}, we can remove the interference component of~\eqref{EQ:SINR_M} similarly to what is commonly done in the literature in successive interference cancellation~\cite{brighente2018power}. We can hence write the \ac{snr} at Bob's side as
\begin{equation}
    \label{EQ:SINR_MHat}
    \hat{\gamma}_{M} = \frac{|h_B|^2 P_t^2}{\sigma_B^2},
\end{equation}
and modify the secrecy capacity in~\eqref{eq:SecCap} as
\begin{equation}
	\hat{C}_{s}= 
	\begin{cases}
		\frac{1}{2} \log_2 \frac{1+\hat{\gamma}_{M}}{1+\gamma_{E}}, \hspace{63pt} \text{if $\gamma_{M} > \gamma_{E}$}, \\
		0, \hspace{110pt} \text{if $\gamma_{M} \leq \gamma_{E}$}.
    \end{cases}
    \label{eq:SecCapMod}
\end{equation}

Thanks to the interference removal capacity of \ac{wbplsec}, Bob's spectral efficiency is now independent from the \ac{ris} phases, and can hence be removed from the optimization problem~\eqref{prob:overall}, which can hence be rewritten as
\begin{subequations}\label{prob:overall2}
\begin{equation}\label{eq:objFun}
    \max_{\bm{\Theta}} \, -\frac{1}{2} \log_2(1+\gamma_E),
\end{equation}
\begin{equation}
    \text{s.t.} \, \theta_k \in [0, 2 \pi], \, \forall k=1, \cdots, K;
\end{equation}
\end{subequations}
where we can rewrite the objective function~\eqref{eq:objFun} as
\begin{equation}\label{eq:newObj}
    \min_{\bm{\Theta}} \, \log_2(1+\gamma_E),
\end{equation}
where we neglected the term $1/2$ as it is constant.

From~\eqref{eq:newObj} we see that our goal is now the minimization of the \ac{sinr} at Eve's side. Considering that the variables of~\eqref{prob:overall2} are the components of matrix $\bm{\Theta}$ uniquely present at the denominator of~\eqref{EQ:SINR_E}, we can rewrite problem~\eqref{prob:overall2} as 

\begin{subequations}\label{prob:overall3}
\begin{equation}\label{eq:objFun2}
    \max_{\bm{\delta}} \, |\bm{h}_{RE}(\delta) \bm{h}_{JR}|^2,
\end{equation}
\begin{equation}
    \text{s.t.} \, \delta_k \in [0, 2 \pi], \, \forall k=1, \cdots, K.
\end{equation}
\end{subequations}

Problem~\eqref{prob:overall3} is non-convex, therefore it is difficult to find an optimal solution. We propose the use of \ac{pso}, a population-based optimization technique, to find the best solution~\cite{kennedy1995particle}. Due to space constraint we do not provide a technical description of \ac{pso}. The interested reader can refer to~\cite{kennedy1995particle, brighente2019location} for more details and examples.

\subsection{Unknown Eve Location}\label{sec:unknown}
In real-life scenarios, the location of Eve can be unknown to the jammer. Therefore, it is not possible to optimally configure the \ac{ris} to convey the jamming signal towards Eve. To circumvent this problem, instead of optimizing the secrecy capacity over a single known Eve location, we define a grid of possible Eve locations distributed over a predefined \textit{area of interest} ($\mathcal{A}$). In this case, the objective function of our optimization problem is given by the area secrecy capacity in~\eqref{eq:areaSec}, and problem~\eqref{prob:overall} can be rewritten as
\begin{subequations}\label{prob:overArea}
\begin{equation}
    \max_{\bm{\Theta}} C_s(\mathcal{A})
\end{equation}
\begin{equation}
    \text{s.t.} \, \theta_k \in [0, 2 \pi], \, \forall k=1, \cdots, K.
\end{equation}
\end{subequations}
In this case, the objective of the optimization problem is to find a configuration of the \ac{ris} that provides the best trade-off among the secrecy capacities obtained evaluating  the possible Eve's locations over $\mathcal{A}$.
As in problem~\eqref{prob:overall}, we set as variable matrix $\bm{\Theta}$ and do not consider power optimization due to the same previously discussed motivations.

Without loss of generality, to simplify problem~\eqref{prob:overArea} from a computational perspective, we replace the continuous area $\mathcal{A}$ with a discrete version considering a gird of points $\{a_1,a_2, \ldots, a_{|\mathcal{A}|}\} \in \mathcal{A}$. We can hence rewrite the area secrecy~\eqref{eq:areaSec} as
\begin{equation}
\begin{split}
    C_s(\mathcal{A}) = & \frac{1}{|\mathcal{A}|}\sum_{a \in \mathcal{A}} \max \{C_{M} - C_{E}(a),0\}   =   \\
    = & \begin{cases}
		\frac{1}{2|\mathcal{A}|}\sum\limits_{a \in \mathcal{A}} \log_2 \frac{1+\gamma_{M}}{1+\gamma_{E}(a)}, \hspace{10pt} \text{if $\gamma_{M} > \gamma_{E}(a)$}, \\
		0, \hspace{94pt} \text{if $\gamma_{M} \leq \gamma_{E}(a)$}.
    \end{cases}
    \end{split}
    \label{eq:discAreaSec}
\end{equation}
Applying the same considerations on the effect of \ac{wbplsec} on Bob's \ac{sinr} and on the dependencies of variables we outlined in Section~\ref{sec:known}, we can rewrite problem~\eqref{prob:overArea} as 
\begin{subequations}\label{prob:overArea2}
\begin{equation}
    \min_{\bm{\Theta}} \sum_{a \in \mathcal{A}} \log_2(1+\gamma_E(a))
\end{equation}
\begin{equation}
    \text{s.t.} \, \theta_k \in [0, 2 \pi], \, \forall k=1, \cdots, K.
\end{equation}
\end{subequations}

The resulting problem~\eqref{prob:overArea2} is non-convex, and it is hence difficult to find an optimal solution. Also in this case, we resort to \ac{pso} to efficiently find the best solution~\cite{kennedy1995particle}.

\section{Simulations Results}
\label{sec:eval}

\begin{figure}[t]
	\centering
	\includegraphics[width=0.83\columnwidth]{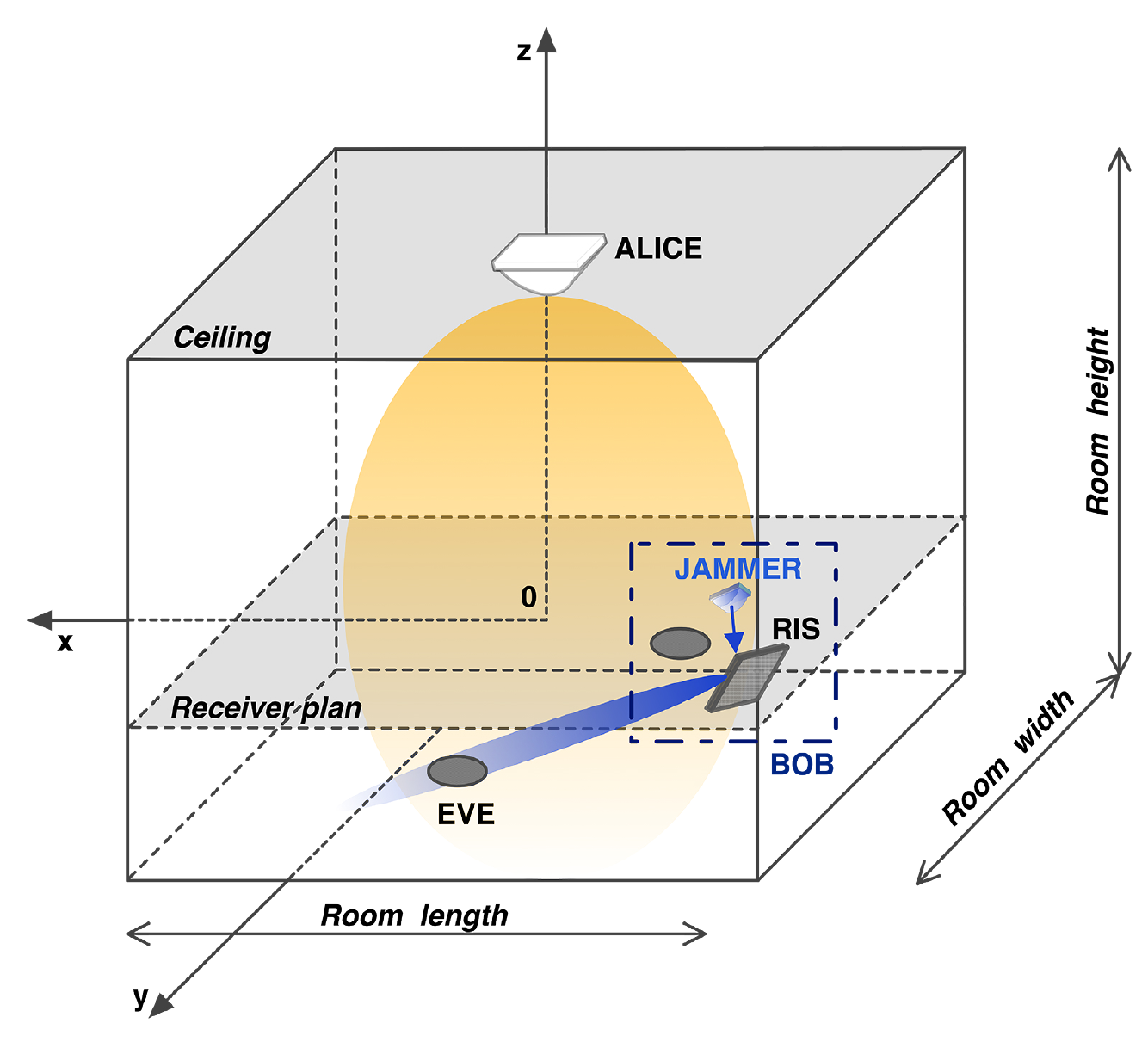}
	\caption{Indoor \ac{vlc} with jammer and \ac{ris}.}
	\label{FIG:ROOM}
	\vspace{-1em}
\end{figure}

Table~\ref{TAB:SIM-PARAM} reports the parameters used for the parametric analysis, including the transmitted power, the jamming intensity, and the orientation of the transmitter and receiver. The \acp{led} used by Alice, Bob, and Eve are identical in their characteristics. For simulations, we assumed Alice was installed on the roof, while Bob and Eve could freely move within the room.
\begin{table}[t]
	\centering
	\captionsetup{justification=centering}
	\caption{Simulation parameters.} \label{TAB:SIM-PARAM}
	\begin{threeparttable}
		\renewcommand{\arraystretch}{1}  
		\begin{tabular}{ll}
			\hline\noalign{\smallskip}
			\bfseries Parameter & \bfseries Value \\
			\hline\noalign{\smallskip}
			
			RIS size ($K$)  & $4,8,16,20,32$ \\
			
			$P_t$  & $1$ $W$    \\
			
			$P_j = P_t (\sfrac{M}{N})$ & ($0.1$, $0.5$) $W$ \\
			
			$\sigma$  &  $10^{-10}$ \\
			
			$A_r$ & $1$ $cm^2$ \\
			
			$\psi_{FOV}$ & $120^\circ$ \\
			
			$\phi_{\frac{1}{2}}$ &  $70^\circ$\\
			
			$\rho$ &  $0.8$ \\
			
			$n$ & $1.5$ \\
			
			$R$  &     $1$ $\sfrac{A}{W}$ \\
			
			Room (length, width, height) & $5$ $\times$ $5$ $\times$ $4$ $m$\\
			
			Alice coordinates ($x$,$y$,$z$) & ($0$,$0$,$2$) $m$\\
			
			Bob coordinates ($x$,$y$,$z$) & ($-1$,$1$,$-0.5$) $m$\\
			
			\hline\noalign{\smallskip}
		\end{tabular}
	\end{threeparttable}
	\vspace{-1em}
\end{table}

To validate our proposed approach, we designed a MATLAB-based simulator to solve problems~\eqref{prob:overall3} and~\eqref{prob:overArea2}, where we considered the indoor scenario with parameters reported in Table~\ref{TAB:SIM-PARAM}. With the architecture proposed (see Figure~\ref{FIG:ROOM}), Bob's jammer is installed close to the \ac{ris} and the jamming channel cannot be described using~\eqref{EQ:VLC_LOS} that is valid only in the far-field region. For this reason, we assume that $P_J$ undergoes a deterministic path-loss attenuation in the near-field region at the \ac{ris}. For both known and unknown Eve's location cases, we fix the location of Alice, Bob, the jammer, and the \ac{ris}, while we consider a grid $\mathcal{N}$ of Eve's locations.
We validate our approach based on two different metrics: the secrecy capacity~\eqref{eq:SecCapMod} achieved at each Eve's location and the outage probability defined as 
\begin{equation}
    P_{\rm out} = \frac{\sum\limits_{n \in \mathcal{N}}\mathbbm{1}\left (\hat{C}_s(n) - T_h \cdot \hat{C}_s \vert_{\rm max} \right  )}{|\mathcal{N}|},
    \label{eq:pOut}
\end{equation}
where $\hat{C}_s \vert_{\rm max} = \max\limits_{n \in \mathcal{N}} \hat{C}_s(n)$, $|\mathcal{N}|$ denotes the cardinality of $\mathcal{N}$, $T_h$ $\in [0, 1]$ is a variable we select to define the secrecy capacity threshold, and $\mathbbm{1}$ is the indicator function
\begin{equation}
    \mathbbm{1}(x) = 
    \begin{cases}
    1, \, \text{if} \, x \geq 0; \\
    0, \, \text{otherwise}.
    \end{cases}
\end{equation}
In other words, the outage probability~\eqref{eq:pOut} provides, given a selected $T_h$, the fraction of points over Eve's grid of location where we achieve a secrecy capacity higher or equal than the $T_h$ percent of the highest achievable secrecy capacity $C_{\rm max}$.

Notice that we use the secrecy capacity~\eqref{eq:SecCapMod} for both known and unknown Eve's location cases. When Eve's location is known, this metric choice is trivial. 
When we solve the optimization problem without knowing Eve's location, the RIS phases direct the jamming to an area where we assume the adversary is. However, this might include the case where the area of interest does not include Eve's actual location. This represents a worst-case scenario, as the hypothesis over the possible Eve's locations is wrong. Keeping the value of the phases fixed, we used the punctual secrecy capacity to calculate the system's performance also when the RIS does not direct the jamming directly to Eve's location.

\begin{figure}[t]
     \centering
     \includegraphics[width=0.9\columnwidth]{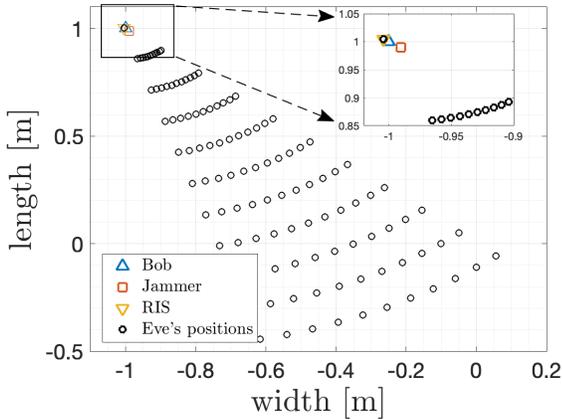}
     \caption{Area of the room where we assume the attacker, i.e. Eve, is positioned.}
     \label{fig:loc}
     \vspace{-2em}
\end{figure}

Figure~\ref{fig:loc} shows the scenario configuration with a view from the above. The \ac{ris} is in the upper left corner and is irradiated by the signal coming from Bob's jammer. We also include the points of Eve's possible locations in the area of interest $\mathcal{A}$, i.e., all the points considered in the area secrecy rate computation.

\begin{figure}[t]
\centering
    \subfigure[center][$K=4$]
    {\centering\includegraphics[width =.48\columnwidth, trim={0.05cm 0.12cm 0.5cm 0.7cm},clip]{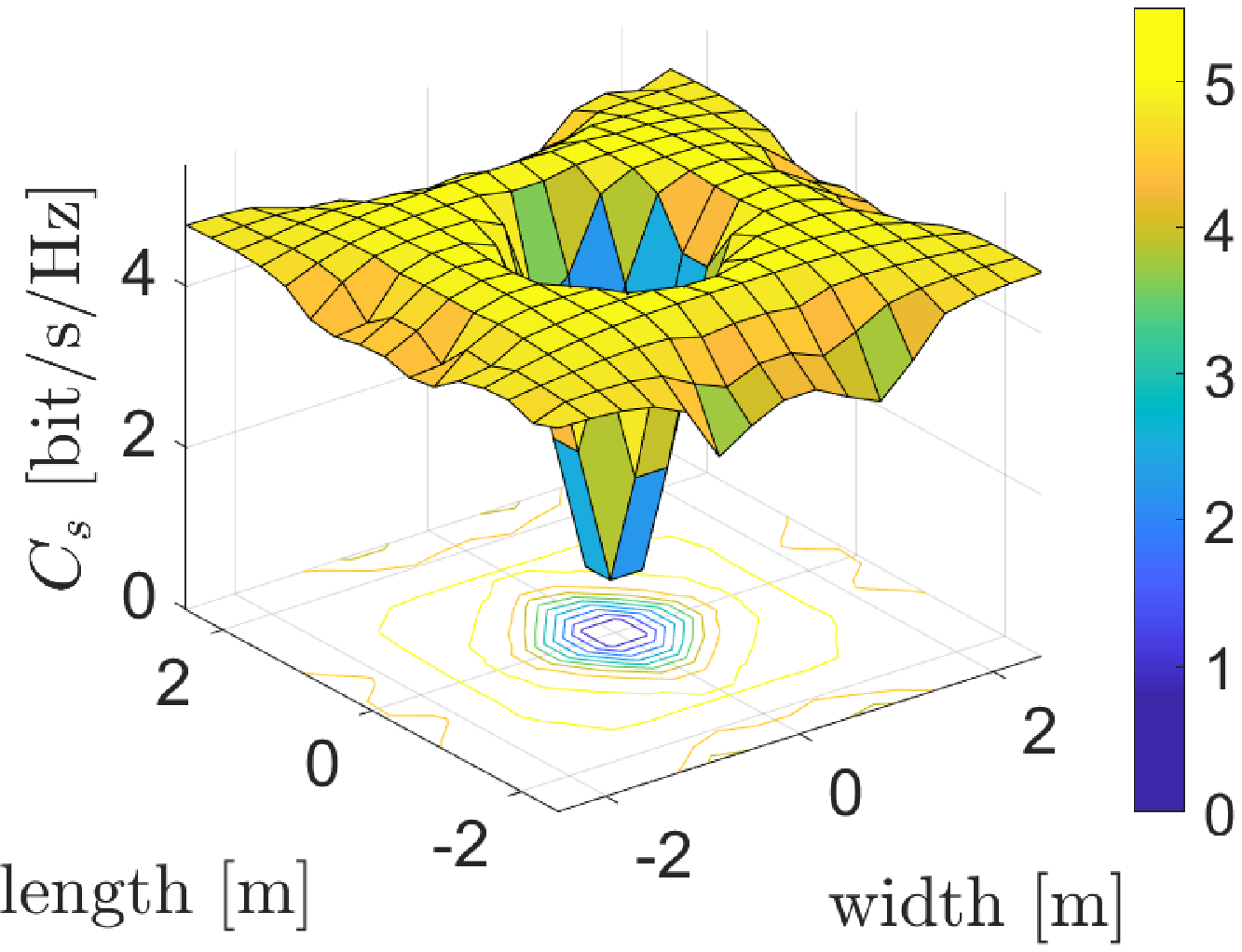}}
    \subfigure[center][$K=8$]{\centering
    \includegraphics[width =.49\columnwidth, trim={0.05cm 0.12cm 0.5cm 0.7cm},clip]{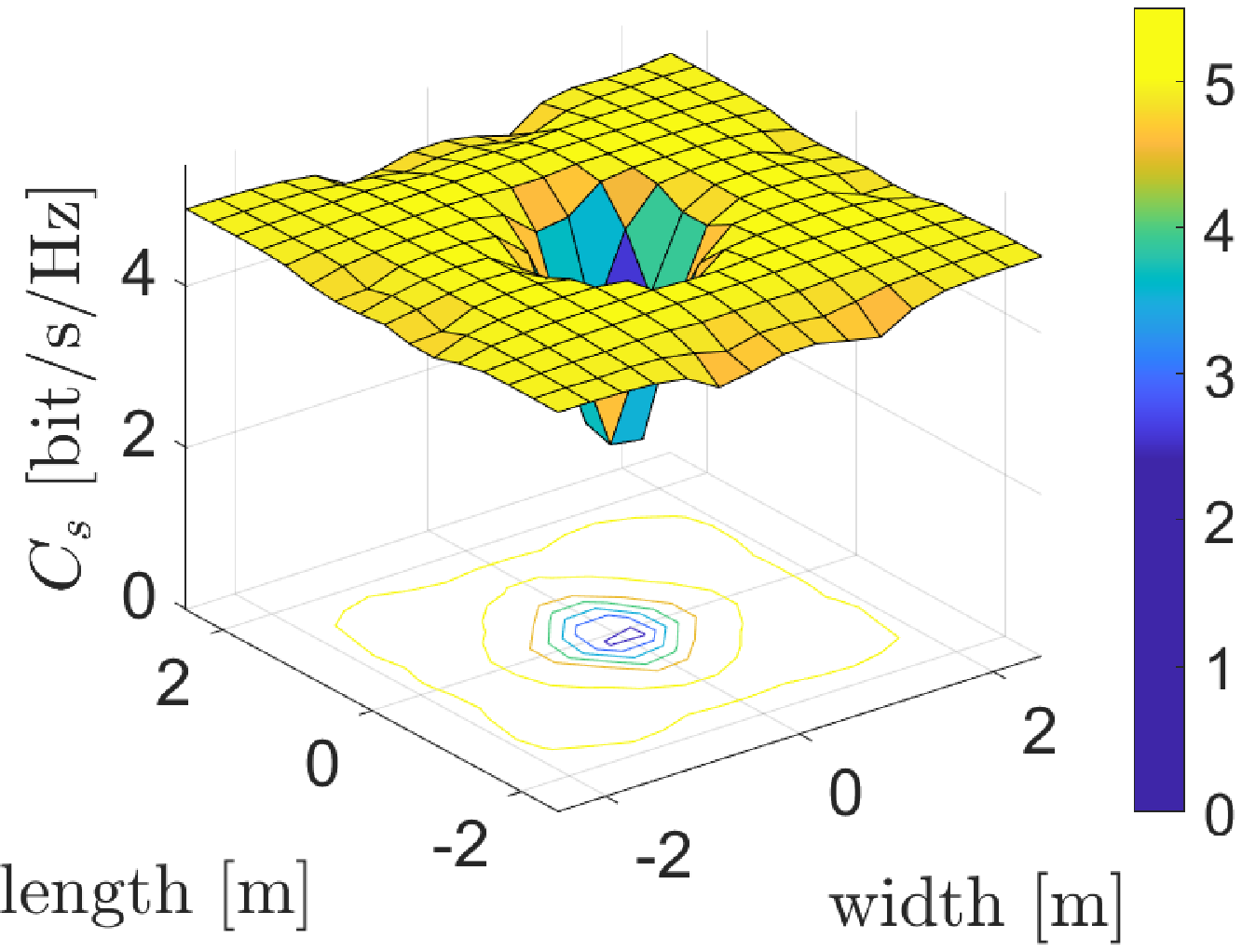}}
    \subfigure[center][$K=16$]{\centering
    \includegraphics[width =.48\columnwidth, trim={0.05cm 0.12cm 0.5cm 0.7cm},clip]{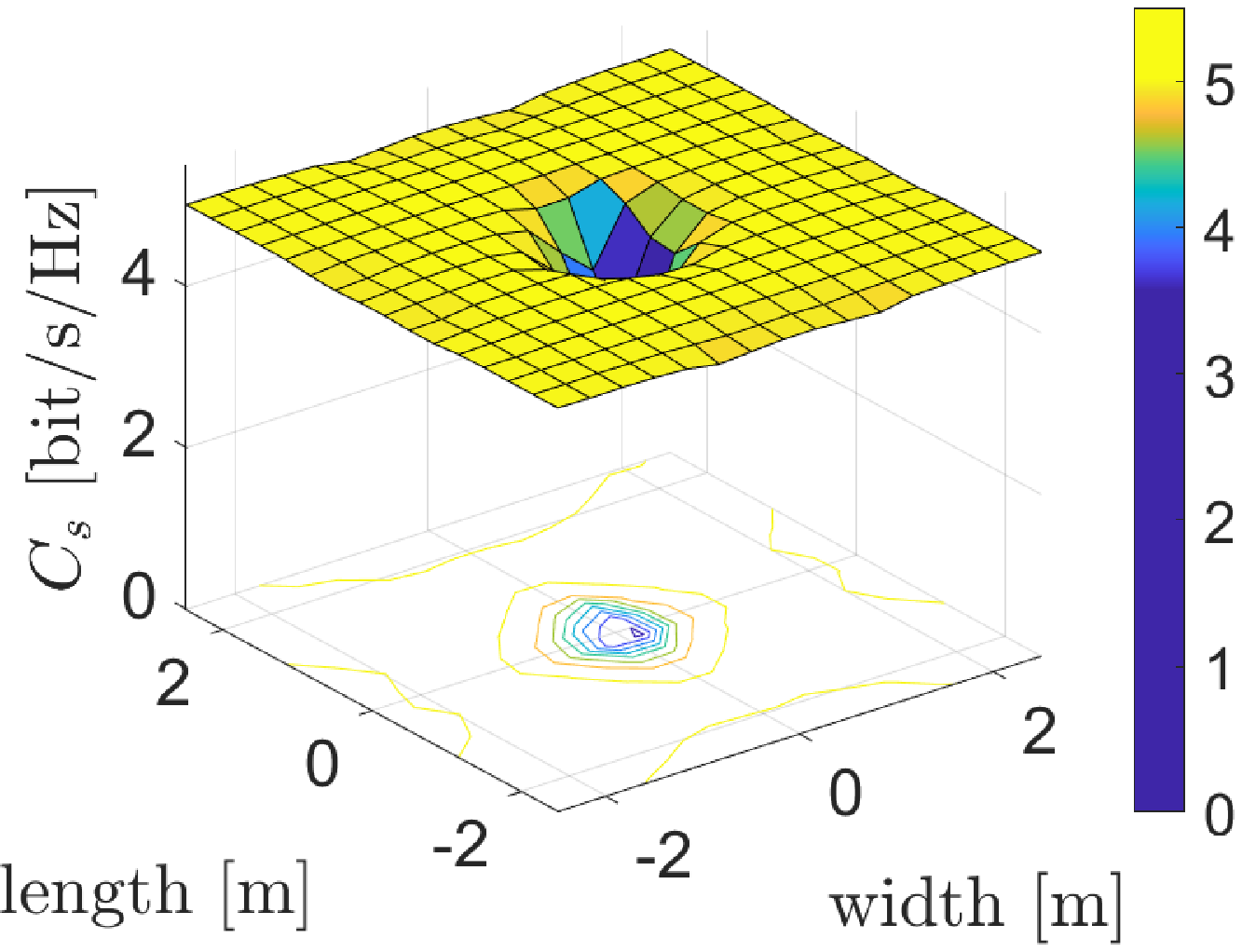}}
    \subfigure[center][$K=32$]{\centering
    \includegraphics[width =.48\columnwidth, trim={0.05cm 0.12cm 0.5cm 0.7cm},clip]{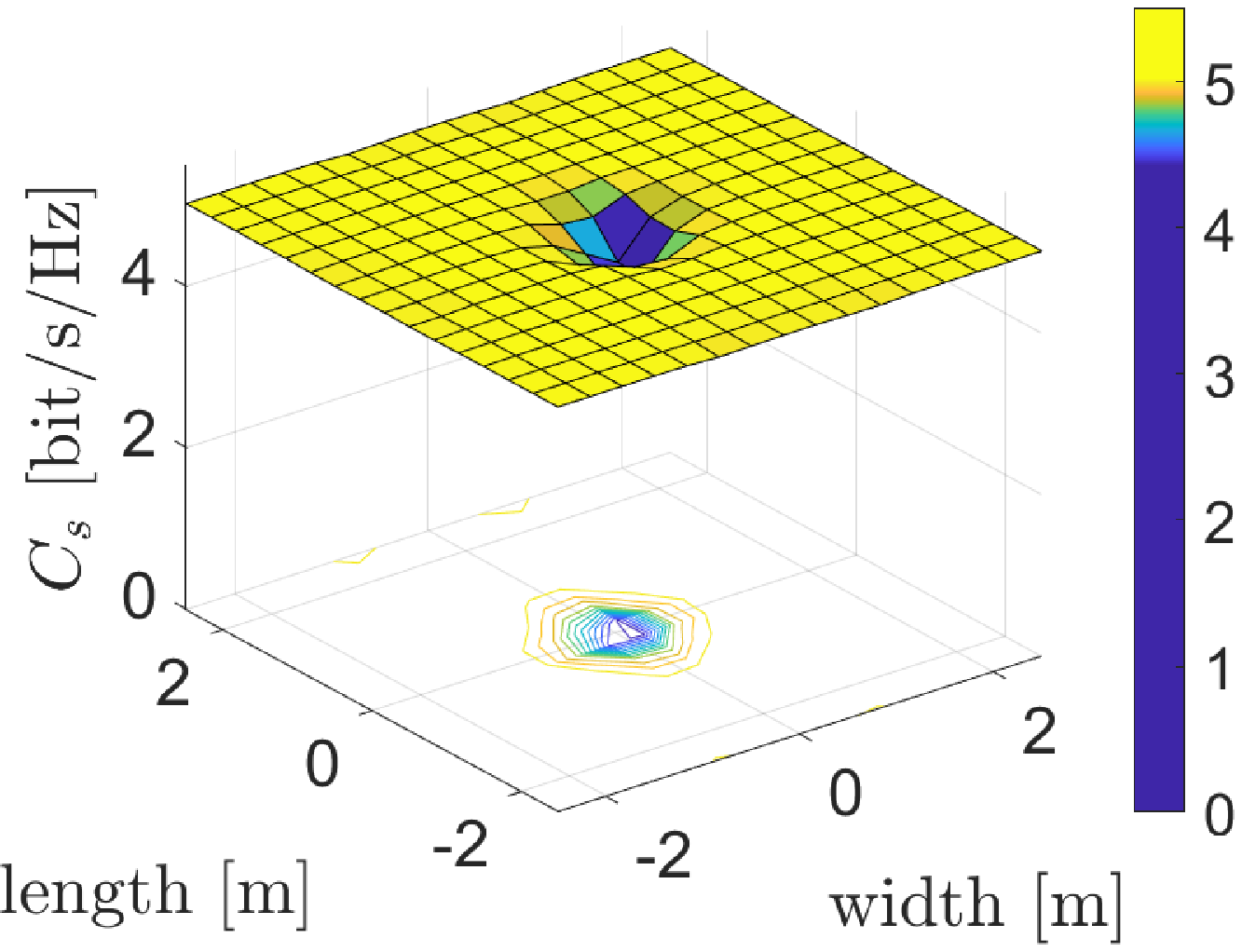}}
 	\caption{Secrecy capacity $C_s$ with known Eve location and $\hat{\gamma}_{M}$~=~$\gamma_E$~=~$15$~dB by varying the number $K$ of \ac{ris} elements.}
 	\label{fig:allKn}
 	\vspace{-1em}
\end{figure}

Figure~\ref{fig:allKn} shows the secrecy rate obtained for different Eve locations around the room when increasing the number $K$ of \ac{ris} elements and considering known Eve locations. We notice that such an increase provides better performance in terms of secrecy rate over the entire area and except around Alice location, the weak spot of \ac{wbplsec}~\cite{2017:soderi_wbplsec_trans}.
\begin{figure}[t]
\centering
    \subfigure[center][$K=4$]
    {\centering\includegraphics[width =.48\columnwidth, trim={0.05cm 0.12cm 0.5cm 0.7cm},clip]{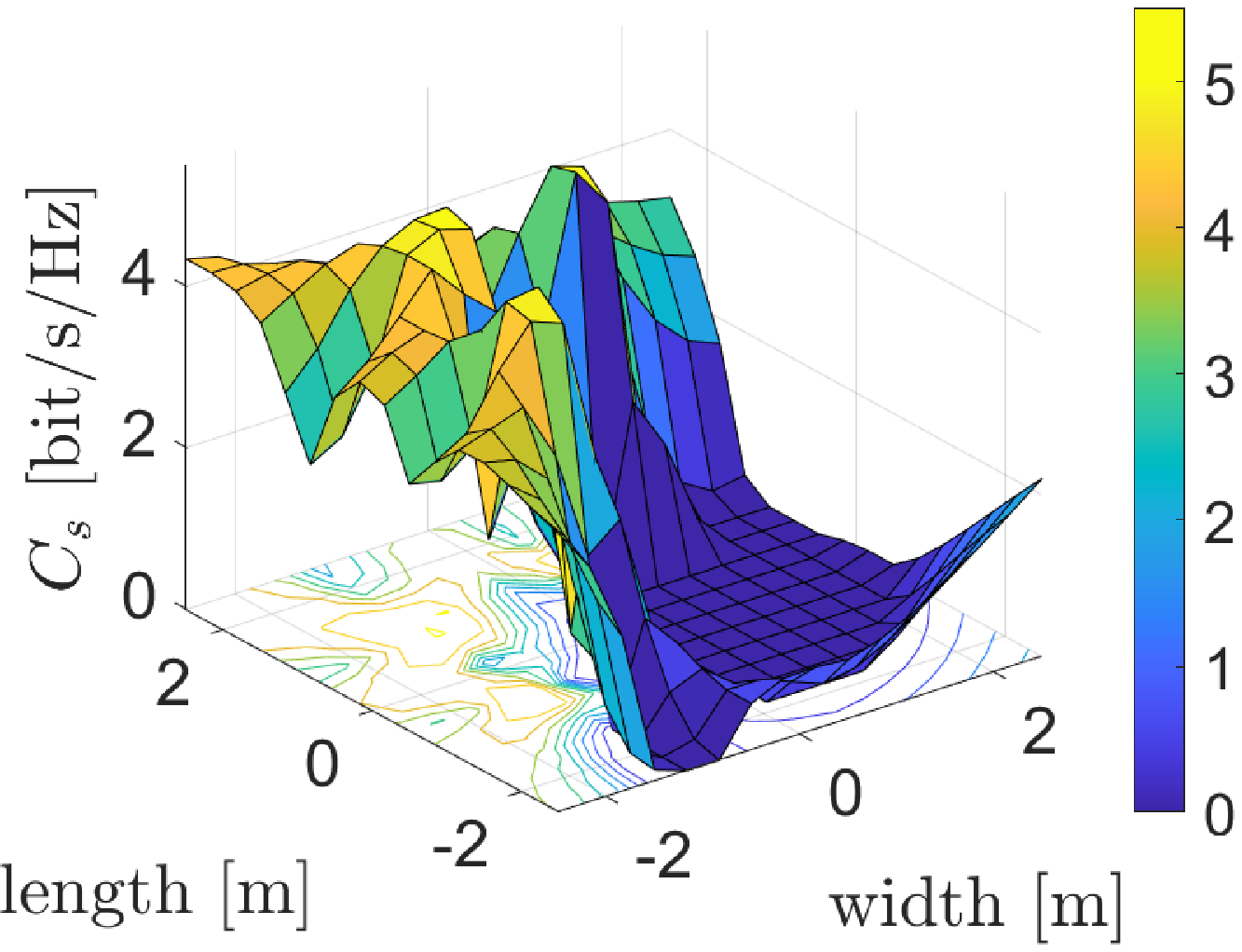}}
    \subfigure[center][$K=8$]{\centering
    \includegraphics[width =.48\columnwidth, trim={0.05cm 0.12cm 0.5cm 0.7cm},clip]{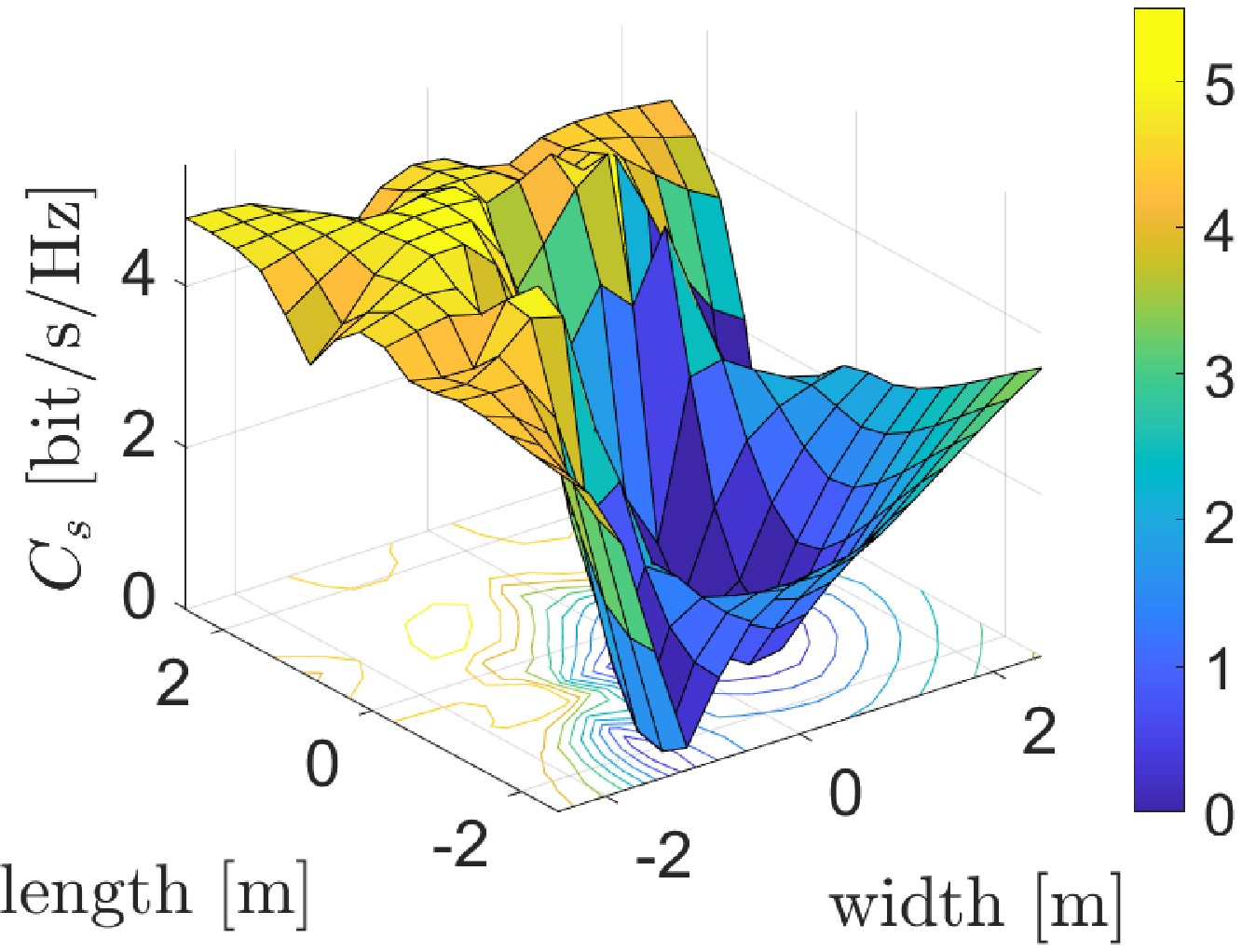}}
    \subfigure[center][$K=16$]{\centering
    \includegraphics[width =.48\columnwidth, trim={0.05cm 0.12cm 0.5cm 0.7cm},clip]{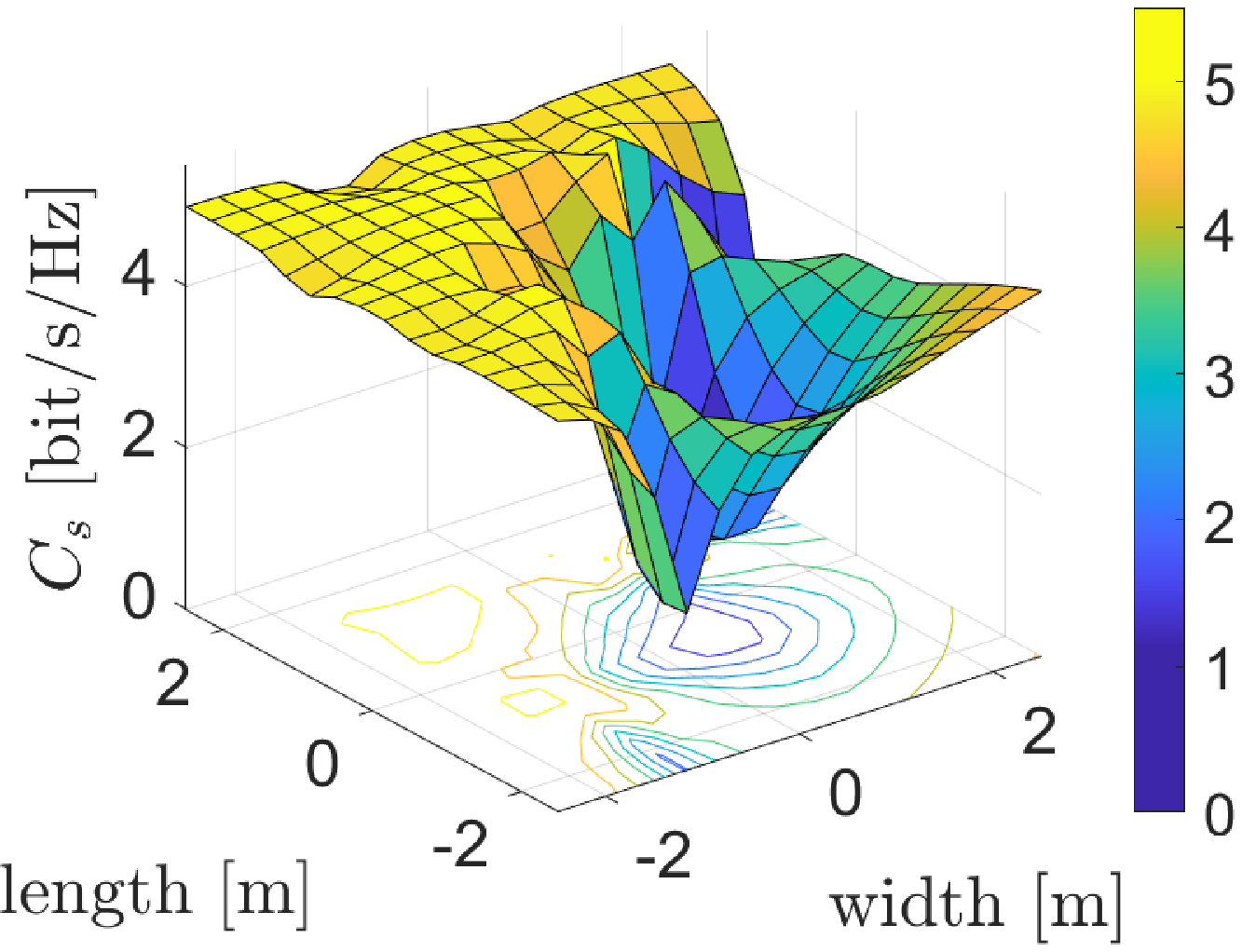}}
    \subfigure[center][$K=32$]{\centering
    \includegraphics[width =.48\columnwidth, trim={0.05cm 0.12cm 0.5cm 0.7cm},clip]{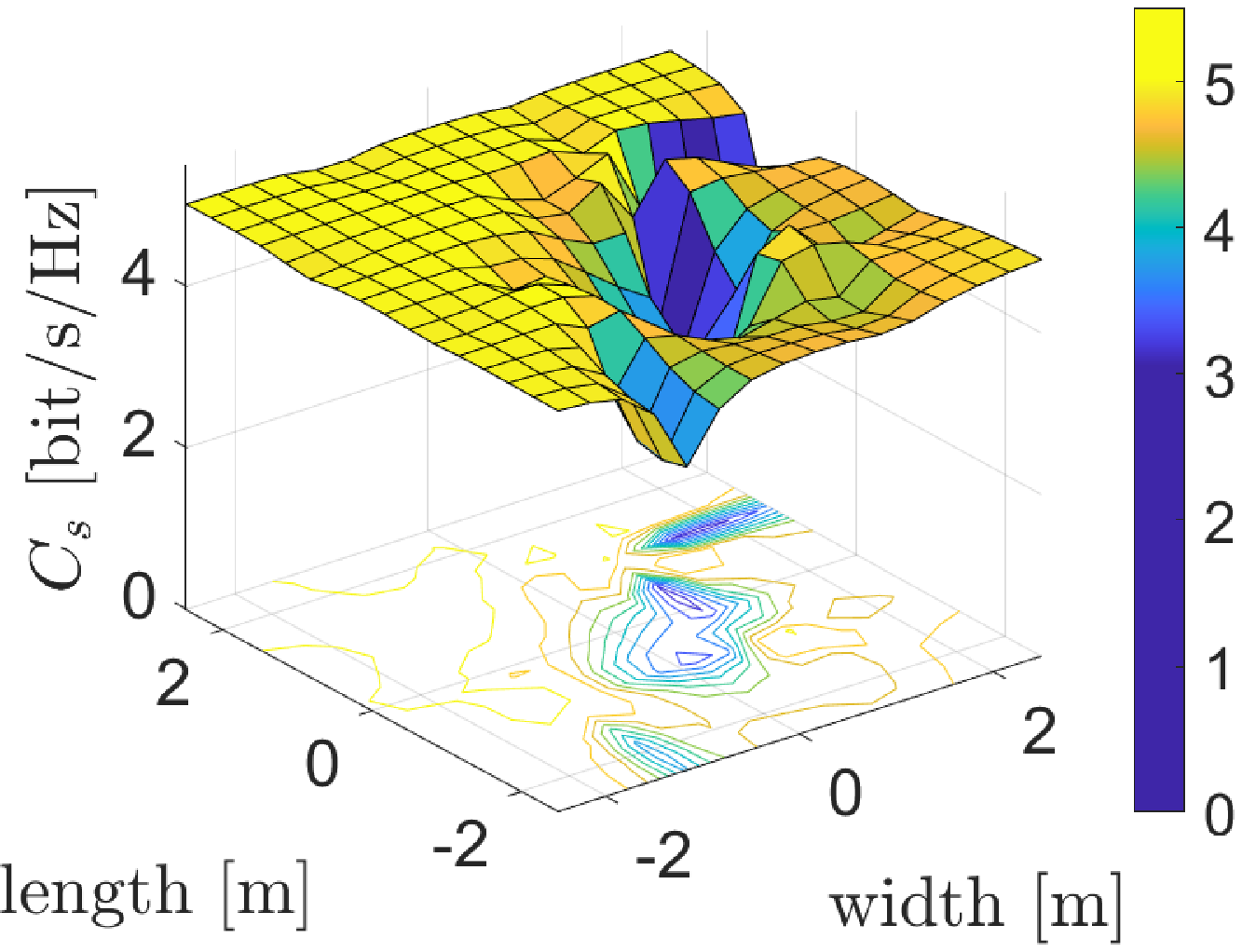}}
 	\caption{Secrecy capacity $C_s$ with unknown Eve location and $\hat{\gamma}_{M}$~=~$\gamma_E$~=~$15$~dB by varying the number $K$ of \ac{ris} elements.}
 	\label{fig:allUnk}
 	\vspace{-1em}
\end{figure}

Figure~\ref{fig:allUnk} shows the secrecy rate obtained for different Eve locations around the room when increasing the number $K$ of \ac{ris} elements and considering unknown Eve locations. As for Figure~\ref{fig:allKn}, we notice that increasing the \ac{ris} size provides performance improval especially around Bob location. Comparing these results with those in Figure~\ref{fig:allKn}, we notice that, thanks to our formulation, removing the hypothesis on known Eve locations dos not worsen the system secrecy.
\begin{figure}
\centering
    \subfigure[center][Known Eve Location]
    {\centering\includegraphics[width =.49\columnwidth]{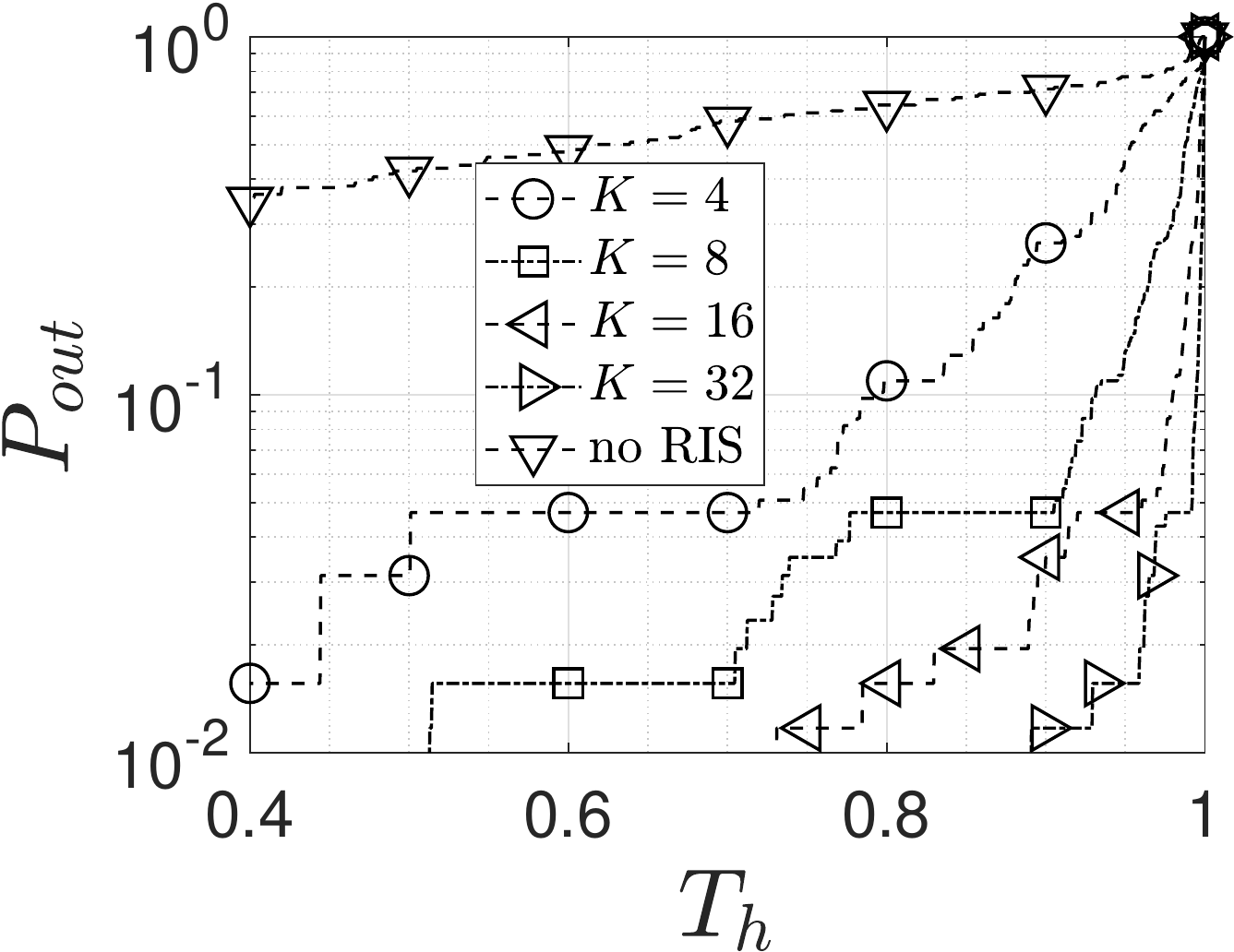}\label{subfig:knPj}}
    \subfigure[center][Unknown Eve Location]{\centering
    \includegraphics[width =.49\columnwidth]{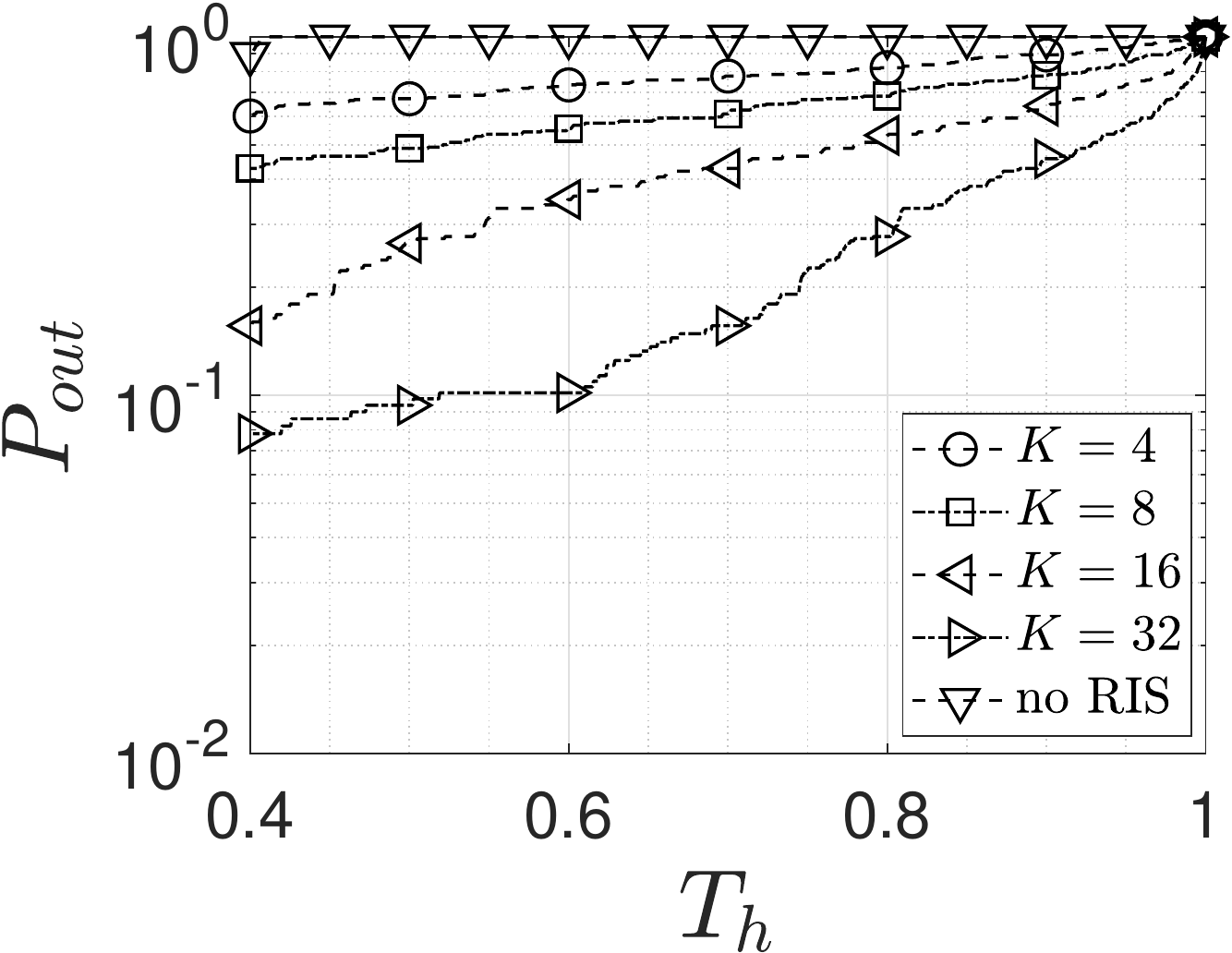}\label{subfig:unknPj}}
 	\caption{$P_{\rm out}$ vs th for different number of \ac{ris} elements. Comparison between known Eve's location (a) and unknown Eve's location (b).}
 	\label{fig:th_2}
 	\vspace{-1em}
\end{figure}

\begin{figure}
\centering
    \subfigure[center][Known Eve Location]
    {\centering\includegraphics[width =.48\columnwidth]{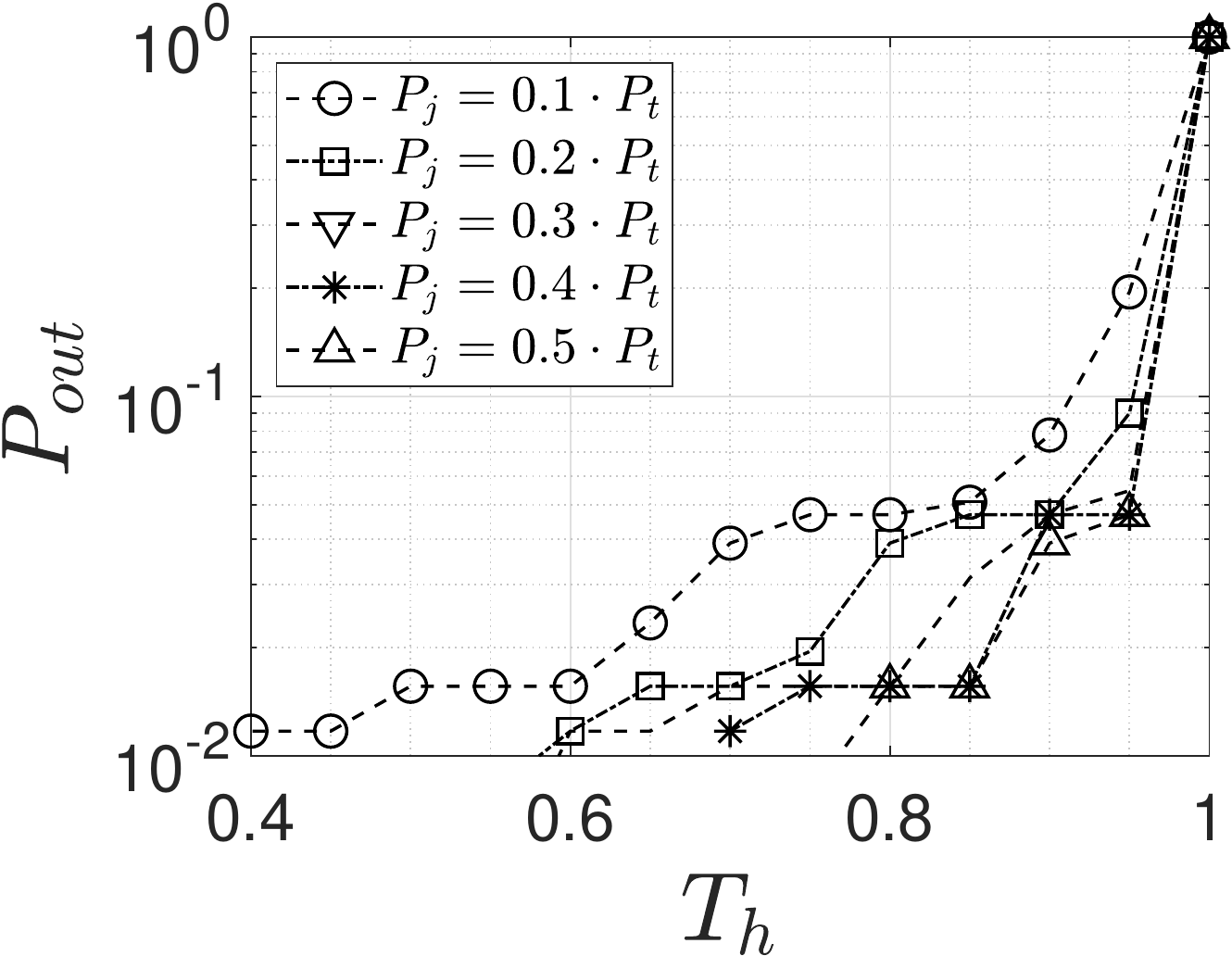}\label{subfig:th1}}
    \subfigure[center][Unknown Eve Location]{\centering
    \includegraphics[width =.48\columnwidth]{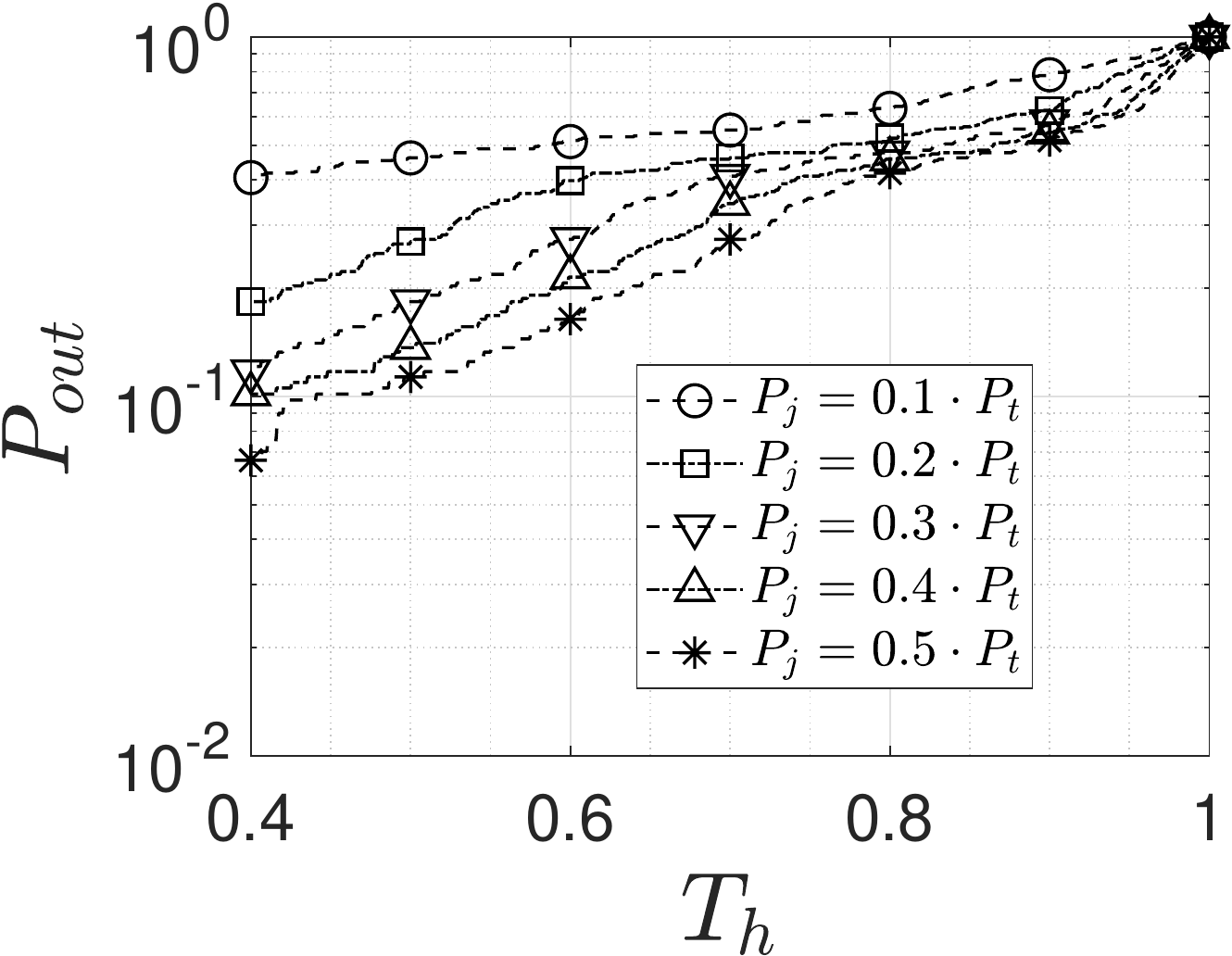}\label{subfig:th2}}
 	\caption{$P_{\rm out}$ vs th for different jamming transmit power. Comparison between known Eve's location (a) and unknown Eve's location (b).}
 	\label{fig:th}
 	\vspace{-1em}
\end{figure}

Figure~\ref{fig:th} shows the outage probability~\eqref{eq:pOut} vs. the threshold $T_h$ obtained without \ac{ris} and for increasing number of \ac{ris} elements. In particular, Figure~\ref{subfig:th1} shows the results obtained with known Eve location, and Figure~\ref{subfig:th2} those obtained with unknown Eve location. In both figures we can see the advantage brought by the introduction of the \ac{ris}. Furthermore, we notice that the outage probability tends to zero when increasing the \ac{ris} size in both known and unknown Eve location cases.

\section{Security Analysis}\label{sec:security}

In this section we analyze the security properties of our schema. In Section~\ref{subsec:threat} we discuss the possible threat model and attacker capabilities. Then, in Section~\ref{subsec:sec_prop} we analyze the security properties which our methodology ensures.

\subsection{Threat Model}\label{subsec:threat}

To assess the robustness of our design, we use a threat model in which an attacker could interfere with Alice and Bob at any moment. In particular, we consider the scenario where an attacker can obtain access to the communication channel during transmission and is therefore capable of both receiving and transmitting. An attacker in this setting could passively eavesdrop or modify the communication.
The most common attacks that an attacker can perform on \ac{vlc} communication can be summarized as follows.
\begin{itemize}
    \item \textbf{Message Injection Attack}: This attack involves sending a malicious and customized message to Bob, for instance, with a malicious command.
    \item\textbf{Replay Attack}: An attacker reuses a previously transmitted and sniffed message in subsequent communications to replicate the legitimate transmission.
    \item\textbf{Message Modification}: Here, the attacker attempts to modify the message during transmission. Unlike the Message Injection Attack, the attacker modifies the message in real-time during a transmission.
    \item\textbf{Eavesdropping}: Attackers use passive sniffing to intercept and collect communication so they can analyze it a second time to compromise future communication.
    \item\textbf{Adversarial Jamming}: Here, the attacker leverages disturbing interference to disrupt the communication and make it no anymore available (\textit{Denial of Service)}.
\end{itemize}

\subsection{Security Properties}\label{subsec:sec_prop}

In the following, we discuss the security features of our approach and we state which of the attacks described in the previous section can be prevented by each security property. Note that these properties are achieved at the physical layer level, allowing subsequent protection to all the above layers.

\parag{Replay attack Resistance} Generally, this feature requires synchronization and nonce exchange between the parties. In our schema, replay attack protection is guaranteed because Bob randomly chose and destroyed a set of bits. Only he knows this information, and the next time a new secret is shared, the jammed bits will be different. In this way, Eve cannot reuse the old messages. This property allows protecting from \textit{Replay} and \textit{Message Injection Attacks}.

\parag{Confidentiality} The message's confidentiality is ensured by the jamming phase. Indeed, only Bob know the jamming points and therefore reconstruct the message. 
Since Bob jams a maximum of $M$ bits out of the total $N$ bits transmitted by Alice, to force the message, the attacker should calculate $2^M$.
This property allows the protection from \textit{Eavesdropping}.

\parag{Integrity} There are two possibilities for an attacker to modify the message: in real-time or in a secondary moment. While the first approach is unfeasible due to the anti-replay property, the second would compromise the watermark, raising errors during the demodulation and watermarking verification. This property prevents \textit{Message Injection} and \textit{Message Modification} attacks. Furthermore, if the attacker performs jamming to perturbate the information, the integrity verification property identifies an unexpected message alteration.  

\parag{Jamming Resistance} Thanks to the watermark, our mechanism can partially mitigate \textit{Adversarial Jamming} attacks. In fact, in the most favorable case, if the adversary would destroy the frames $N_W$ frames dedicated to transmitting the watermark, Bob can still reconstruct the original message using the \ac{wbplsec} algorithm.
Instead, if Eve jams other frames, our mechanism will still suffer from this type of attack.

\section{Conclusion}\label{sec:conclusion}

\ac{vlc} is considered a key technology for future wireless communications and the introduction of \ac{ris} technology can improve the quality of the received beam. 
To the best of our knowledge, we propose for the first time in literature that using \ac{ris} on \ac{vlc} improves the effectiveness of jamming in \ac{wbplsec} without any assumptions about the attacker's location.
We evaluated the robustness of our approach in an indoor wireless communications scenario, where the \ac{vlc} transmitter on the ceiling can move unconstrained further down. We considered a \ac{vlc} channel model with only the \ac{los} component for numerical simulations.

Our results show that the introduction of \ac{ris} technology extends the area where secure communication occurs and that by increasing the number of \ac{ris} elements the outage probability decreases.

Unfortunately our proposal suffers from some limitations.
First, the channel is simulated based on a widely accepted geometrical model, which can be limited compared to a real device implementation. However, currently no works are using real \ac{ris} hardware for \ac{vlc} system. Second, since the problem is non-convex, \ac{pso} may not identify the optimal solution. For this reason we will investigate other approaches in future works.

\balance
\bibliographystyle{IEEEtran}
\bibliography{bib}

\begin{thebibliography}{10}
\providecommand{\url}[1]{#1}
\csname url@samestyle\endcsname
\providecommand{\newblock}{\relax}
\providecommand{\bibinfo}[2]{#2}
\providecommand{\BIBentrySTDinterwordspacing}{\spaceskip=0pt\relax}
\providecommand{\BIBentryALTinterwordstretchfactor}{4}
\providecommand{\BIBentryALTinterwordspacing}{\spaceskip=\fontdimen2\font plus
\BIBentryALTinterwordstretchfactor\fontdimen3\font minus
  \fontdimen4\font\relax}
\providecommand{\BIBforeignlanguage}[2]{{%
\expandafter\ifx\csname l@#1\endcsname\relax
\typeout{** WARNING: IEEEtran.bst: No hyphenation pattern has been}%
\typeout{** loaded for the language `#1'. Using the pattern for}%
\typeout{** the default language instead.}%
\else
\language=\csname l@#1\endcsname
\fi
#2}}
\providecommand{\BIBdecl}{\relax}
\BIBdecl

\bibitem{web:ciscoreport}
\BIBentryALTinterwordspacing
Cisco. (2021) ``{VNI Complete Forecast Highlights}". [Online]. Available:
  \url{https://www.cisco.com/c/dam/m/en\_us/solutions/service-provider\/vni-forecast-highlights/pdf/Global\_2021\_Forecast\_Highlights.pdf}
\BIBentrySTDinterwordspacing

\bibitem{tsonev2015towards}
D.~Tsonev, S.~Videv, and H.~Haas, ``Towards a 100 gb/s visible light wireless
  access network,'' \emph{Optics express}, pp. 1627--1637, 2015.

\bibitem{tiwari2015smart}
S.~V. Tiwari, A.~Sewaiwar, and Y.-H. Chung, ``Smart home technologies using
  visible light communication,'' in \emph{2015 IEEE International Conference on
  Consumer Electronics (ICCE)}.\hskip 1em plus 0.5em minus 0.4em\relax IEEE,
  2015, pp. 379--380.

\bibitem{dahri2019experimental}
F.~A. Dahri, H.~B. Mangrio, A.~Baqai, and F.~A. Umrani, ``Experimental
  evaluation of intelligent transport system with vlc vehicle-to-vehicle
  communication,'' \emph{Wireless Personal Communications}, pp. 1885--1896,
  2019.

\bibitem{acar2017comparing}
Y.~Acar, M.~Backes, S.~Fahl, S.~Garfinkel, D.~Kim, M.~L. Mazurek, and
  C.~Stransky, ``Comparing the usability of cryptographic apis,'' in \emph{IEEE
  Symposium on Security and Privacy (SP)}.\hskip 1em plus 0.5em minus
  0.4em\relax IEEE, 2017, pp. 154--171.

\bibitem{ndjiongue1999visible}
A.~R. Ndjiongue, H.~C. Ferreira, and T.~M. Ngatched, ``Visible light
  communications (vlc) technology,'' \emph{Wiley Encyclopedia of Electrical and
  Electronics Engineering}, pp. 1--15, 1999.

\bibitem{jovicic2013visible}
A.~Jovicic, J.~Li, and T.~Richardson, ``Visible light communication:
  opportunities, challenges and the path to market,'' \emph{IEEE Communications
  Magazine}, vol.~51, no.~12, pp. 26--32, 2013.

\bibitem{rajagopal2012ieee}
S.~Rajagopal, R.~D. Roberts, and S.-K. Lim, ``Ieee 802.15. 7 visible light
  communication: modulation schemes and dimming support,'' \emph{IEEE
  Communications Magazine}, vol.~50, no.~3, pp. 72--82, 2012.

\bibitem{tanaka2000wireless}
Y.~Tanaka, S.~Haruyama, and M.~Nakagawa, ``Wireless optical transmissions with
  white colored led for wireless home links,'' in \emph{11th IEEE International
  Symposium on Personal Indoor and Mobile Radio Communications. PIMRC 2000.
  Proceedings (Cat. No. 00TH8525)}, vol.~2.\hskip 1em plus 0.5em minus
  0.4em\relax IEEE, 2000, pp. 1325--1329.

\bibitem{komar2017electrically}
A.~Komar, Z.~Fang, J.~Bohn, J.~Sautter, M.~Decker, A.~Miroshnichenko,
  T.~Pertsch, I.~Brener, Y.~S. Kivshar, I.~Staude \emph{et~al.}, ``Electrically
  tunable all-dielectric optical metasurfaces based on liquid crystals,''
  \emph{Applied Physics Letters}, vol. 110, no.~7, p. 071109, 2017.

\bibitem{9474926}
A.~R. Ndjiongue, T.~M.~N. Ngatched, O.~A. Dobre, and H.~Haas, ``Toward the use
  of re-configurable intelligent surfaces in vlc systems: Beam steering,''
  \emph{IEEE Wireless Communications}, pp. 156--162, 2021.

\bibitem{mursia2021rise}
P.~Mursia, F.~Devoti, V.~Sciancalepore, and X.~Costa-P{\'e}rez, ``Rise of
  flight: Ris-empowered uav communications for robust and reliable
  air-to-ground networks,'' \emph{arXiv preprint arXiv:2105.03636}, 2021.

\bibitem{ozcan2021reconfigurable}
Y.~U. Ozcan, O.~Ozdemir, and G.~K. Kurt, ``Reconfigurable intelligent surfaces
  for the connectivity of autonomous vehicles,'' \emph{IEEE Transactions on
  Vehicular Technology}, vol.~70, no.~3, pp. 2508--2513, 2021.

\bibitem{BLINOWSKI2019246}
\BIBentryALTinterwordspacing
G.~Blinowski, ``Security of visible light communication systems—a survey,''
  \emph{Physical Communication}, vol.~34, pp. 246--260, 2019. [Online].
  Available:
  \url{https://www.sciencedirect.com/science/article/pii/S1874490718304786}
\BIBentrySTDinterwordspacing

\bibitem{Ndjiongue:21}
\BIBentryALTinterwordspacing
A.~R. Ndjiongue, T.~M.~N. Ngatched, O.~A. Dobre, and H.~Haas, ``Re-configurable
  intelligent surface-based vlc receivers using tunable liquid-crystals: The
  concept,'' \emph{J. Lightwave Technol.}, vol.~39, no.~10, pp. 3193--3200, May
  2021. [Online]. Available:
  \url{http://www.osapublishing.org/jlt/abstract.cfm?URI=jlt-39-10-3193}
\BIBentrySTDinterwordspacing

\bibitem{2021:soderi_pls_rgb_vlc}
S.~Soderi and R.~De~Nicola, ``{6G Networks Physical Layer Security Using RGB
  Visible Light Communications},'' \emph{IEEE Access}, vol.~10, pp. 5482--5496,
  2022.

\bibitem{1997:vlc_channel_barry}
J.~M. {Kahn} and J.~R. {Barry}, ``Wireless infrared communications,''
  \emph{Proceedings of the IEEE}, vol.~85, no.~2, pp. 265--298, Feb 1997.

\bibitem{aboagye2021intelligent}
S.~Aboagye, T.~M. Ngatched, O.~A. Dobre, and A.~R. Ndjiongue, ``Intelligent
  reflecting surface-aided indoor visible light communication systems,''
  \emph{IEEE Communications Letters}, 2021.

\bibitem{2017:soderi_wbplsec_trans}
S.~Soderi, L.~Mucchi, M.~H{\"a}m{\"a}l{\"a}inen, A.~Piva, and J.~H. Iinatti,
  ``{Physical layer security based on spread-spectrum watermarking and jamming
  receiver.}'' \emph{Trans. Emerging Telecommunications Technologies}, vol.~28,
  no.~7, 2017.

\bibitem{2019:soderi_acoustic_journal}
S.~Soderi, ``{Acoustic-Based Security: A Key Enabling Technology for Wireless
  Sensor Network},'' \emph{{International Journal of Wireless Information
  Networks}}, vol.~27, no.~1, pp. 45--59, nov 2019.

\bibitem{2000:Proakis_digital_4th}
\BIBentryALTinterwordspacing
J.~G. Proakis, \emph{{Digital communications}}, 4th~ed.\hskip 1em plus 0.5em
  minus 0.4em\relax Boston: McGraw-Hill, 2000, accessed: 2022-05-05. [Online].
  Available: \url{http://www.loc.gov/catdir/description/mh021/00025305.html}
\BIBentrySTDinterwordspacing

\bibitem{akram2017camera}
M.~Akram, L.~Aravinda, M.~Munaweera, G.~Godaliyadda, and M.~Ekanayake, ``Camera
  based visible light communication system for underwater applications,'' in
  \emph{2017 IEEE International Conference on Industrial and Information
  Systems (ICIIS)}.\hskip 1em plus 0.5em minus 0.4em\relax IEEE, 2017, pp.
  1--6.

\bibitem{arfaoui2020physical}
M.~A. Arfaoui, M.~D. Soltani, I.~Tavakkolnia, A.~Ghrayeb, M.~Safari, C.~M.
  Assi, and H.~Haas, ``Physical layer security for visible light communication
  systems: A survey,'' \emph{IEEE Communications Surveys \& Tutorials},
  vol.~22, no.~3, pp. 1887--1908, 2020.

\bibitem{arfaoui2018secrecy}
M.~A. Arfaoui, A.~Ghrayeb, and C.~M. Assi, ``Secrecy performance of multi-user
  miso vlc broadcast channels with confidential messages,'' \emph{IEEE
  Transactions on Wireless Communications}, vol.~17, no.~11, pp. 7789--7800,
  2018.

\bibitem{chen2017physical}
Z.~Chen and H.~Haas, ``Physical layer security for optical attocell networks,''
  in \emph{2017 IEEE International Conference on Communications (ICC)}.\hskip
  1em plus 0.5em minus 0.4em\relax IEEE, 2017, pp. 1--6.

\bibitem{wang2019enhancing}
F.~Wang, R.~Li, J.~Zhang, S.~Shi, and C.~Liu, ``Enhancing the secrecy
  performance of the spatial modulation aided vlc systems with optical
  jamming,'' \emph{Signal Processing}, vol. 157, pp. 288--302, 2019.

\bibitem{panayirci2020physical}
E.~Panayirci, A.~Yesilkaya, T.~Cogalan, H.~V. Poor, and H.~Haas,
  ``Physical-layer security with optical generalized space shift keying,''
  \emph{IEEE Transactions on Communications}, pp. 3042--3056, 2020.

\bibitem{yang2018mapping}
Y.~Yang and M.~Guizani, ``Mapping-varied spatial modulation for physical layer
  security: Transmission strategy and secrecy rate,'' \emph{IEEE Journal on
  Selected Areas in Communications}, pp. 877--889, 2018.

\bibitem{jiang2017secrecy}
X.-Q. Jiang, M.~Wen, H.~Hai, J.~Li, and S.~Kim, ``Secrecy-enhancing scheme for
  spatial modulation,'' \emph{IEEE Communications Letters}, vol.~22, no.~3, pp.
  550--553, 2017.

\bibitem{abumarshoud2021lifi}
H.~Abumarshoud, L.~Mohjazi, O.~A. Dobre, M.~Di~Renzo, M.~A. Imran, and H.~Haas,
  ``{LiFi Through Reconfigurable Intelligent Surfaces: A New Frontier for
  6G?}'' \emph{arXiv preprint arXiv:2104.02390}, 2021.

\bibitem{li2021robust}
S.~Li, B.~Duo, M.~Di~Renzo, M.~Tao, and X.~Yuan, ``Robust secure uav
  communications with the aid of reconfigurable intelligent surfaces,''
  \emph{IEEE Transactions on Wireless Communications}, 2021.

\bibitem{1978:nonDegradedChannel}
I.~Csiszar and J.~Korner, ``{Broadcast channels with confidential messages},''
  \emph{IEEE Transactions on Information Theory}, vol.~24, no.~3, pp. 339--348,
  May 1978.

\bibitem{2006:Barros_secrecy_wireless_ch}
J.~Barros and M.~R.~D. Rodrigues, ``{Secrecy Capacity of Wireless Channels},''
  in \emph{{2006 IEEE International Symposium on Information Theory}}, July
  2006, pp. 356--360.

\bibitem{uavRis}
\BIBentryALTinterwordspacing
A.~Brighente, M.~Conti, H.~Idriss, and S.~Tomasin, ``Unmanned aerial vehicles
  meet reflective intelligent surfaces to improve coverage and secrecy,'' 2022.
  [Online]. Available: \url{https://arxiv.org/abs/2205.02506}
\BIBentrySTDinterwordspacing

\bibitem{boyd2004convex}
S.~Boyd, S.~P. Boyd, and L.~Vandenberghe, \emph{Convex optimization}.\hskip 1em
  plus 0.5em minus 0.4em\relax Cambridge university press, 2004.

\bibitem{brighente2018power}
A.~Brighente and S.~Tomasin, ``Power allocation for non-orthogonal millimeter
  wave systems with mixed traffic,'' \emph{IEEE Transactions on Wireless
  Communications}, vol.~18, no.~1, pp. 432--443, 2018.

\bibitem{kennedy1995particle}
J.~Kennedy and R.~Eberhart, ``Particle swarm optimization,'' in
  \emph{Proceedings of ICNN'95-international conference on neural networks},
  vol.~4.\hskip 1em plus 0.5em minus 0.4em\relax IEEE, 1995, pp. 1942--1948.

\bibitem{brighente2019location}
A.~Brighente, F.~Formaggio, M.~Centenaro, G.~M. Di~Nunzio, and S.~Tomasin,
  ``Location-verification and network planning via machine learning
  approaches,'' in \emph{2019 International Symposium on Modeling and
  Optimization in Mobile, Ad Hoc, and Wireless Networks (WiOPT)}.\hskip 1em
  plus 0.5em minus 0.4em\relax IEEE, 2019, pp. 1--7.

\end{thebibliography}

\end{document}